\documentclass[mnsc,nonblindrev]{informs3} %

\usepackage{my_style}

\usepackage{hyperref}
\hypersetup{
    colorlinks=true,
    linkcolor=blue,
    citecolor=blue,
    filecolor=magenta,      
    urlcolor=cyan,
    pdftitle={Measuring Strategization in Recommendation},
    pdfpagemode=FullScreen,
    }

\usepackage[capitalise]{cleveref}
\crefname{hypothesis}{Hypothesis}{Hypotheses}

\OneAndAHalfSpacedXI

\usepackage{natbib}
 \bibpunct[, ]{(}{)}{,}{a}{}{,}%
 \def\bibfont{\small}%
\usepackage{booktabs}

\TheoremsNumberedThrough     %
\ECRepeatTheorems

\EquationsNumberedThrough    %

\MANUSCRIPTNO{}

\renewcommand{\theARTICLETOP}{}
\renewcommand{\theLRHSecondLine}{}
\renewcommand{\theRRHSecondLine}{}

\begin{document}

\RUNAUTHOR{Cen, Ilyas et al.}

\RUNTITLE{Measuring Strategization in Recommendation}

\TITLE{Measuring Strategization in Recommendation: Users Adapt Their Behavior to Shape Future Content}

\ARTICLEAUTHORS{%
\AUTHOR{Sarah H. Cen}
\AFF{Department of Electrical Engineering and Computer Science, Massachusetts Institute of Technology, shcen@mit.edu}
\AUTHOR{Andrew Ilyas}
\AFF{Department of Electrical Engineering and Computer Science, Massachusetts Institute of Technology, \EMAIL{ailyas@mit.edu}} %
\AUTHOR{Jennifer Allen}
\AFF{ Sloan School of Management, Massachusetts Institute of Technology, jnallen@mit.edu}
\AUTHOR{Hannah Li}
\AFF{Decision, Risk, and Operations Division, Columbia University, hannah.li@columbia.edu}
\AUTHOR{Aleksander M\k{a}dry}
\AFF{Department of Electrical Engineering and Computer Science, Massachusetts Institute of Technology, madry@mit.edu}
} %

\ABSTRACT{%
     Most modern recommendation algorithms are data-driven: they generate
personalized recommendations by observing users' past behaviors. A common
assumption in recommendation is that how a user interacts with a piece of
content (e.g., whether they choose to ``like'' it) is a reflection of the
content, but \emph{not} of the algorithm that generated it. Although this
assumption is convenient, it fails to capture user strategization: that users
may attempt to shape their future recommendations by adapting their behavior to
the recommendation algorithm. In this work, we test for user strategization by
conducting a lab experiment and survey. 
To capture strategization, 
we adopt a model in which strategic users 
select their engagement behavior based not only on the content, but also on how their behavior affects downstream recommendations. 
Using a custom music player that we built, we study how users respond to different \emph{information} about their recommendation algorithm as well as to different \emph{incentives} about how their actions affect downstream outcomes. We find strong evidence of strategization across outcome metrics, 
including participants' dwell time and use of ``likes.'' 
For example, 
participants who are told that the algorithm mainly pays attention to ``likes'' and ``dislikes'' use those functions 1.9x more than participants told that the algorithm mainly pays attention to dwell time.
A close analysis of participant behavior (e.g., in response to our incentive conditions) rules out experimenter demand as the main driver of these trends. 
Further, in our post-experiment survey, nearly half of participants self-report strategizing ``in the wild,''
with some stating that they ignore content they actually like to avoid over-recommendation of that content in the future. 
Together, our findings suggest that user strategization is common and that platforms cannot ignore the effect of their recommendation algorithms on user behavior.

}%

\maketitle

\section{Introduction} \label{sec:intro}

Platforms like TikTok, Netflix, and Amazon attract and retain users by 
tailoring content (e.g., videos, shows, and products) to each user's interests using \emph{recommendation algorithms}.
While many recommendation algorithms exist,
all of them are trained on \emph{past user behavior}. 
For instance,
Netflix generates recommendations based on each user's watch and rating history.

Recommendation algorithms typically rely on an assumption that user behavior is \emph{exogenous}:
how a user reacts to a recommendation depends on that recommendation alone, 
and \emph{not} on the algorithm that generates it \citep{adomavicius2005toward,ricci2010introduction}. 
This assumption implies, for example, that a user will ``like'' a video with the same probability irrespective of the recommendation algorithm that produces it. 
In other words, 
a user's revealed preferences (implied by their engagement behavior) remain \emph{consistent} across recommendation algorithms as long as their true preferences (their unknown utility function) remain the same.

What this exogeneity assumption fails to capture is \emph{strategic} behavior: 
users may react to recommendations in a way that not only depends on the recommendation, 
but also on their perception of the algorithm itself. 
More precisely, 
strategic behavior implies that users may attempt to shape their future recommendations by adapting their revealed preferences to their recommendation algorithm, 
even if their true preferences do not change. 
For example, a TikTok user might ``heart'' a video not because they enjoy it, 
but because they like the creator and believe TikTok's algorithm will recommend more content from creators they ``heart'' in the future. 
A Spotify user might choose to ignore a ``guilty pleasure'' song that they actually like because they are worried Spotify's algorithm will recommend too many similar songs later on. 
In these examples,
the true, unknown utility that the user receives from each recommendation does not change across algorithms, but the user's behavior may. 
If the user is aware that their actions serve as training data for future recommendations, they may adjust their actions to influence their downstream outcomes.

A recent paper by \citep{cen2023trust}
posits that users do indeed strategize, modeling the interactions between users and platforms as a repeated two-player game. They argue that determining whether users strategize is important because strategization implies that behavioral data gathered under one algorithm is not ``representative'' of the user’s behavior in general. When a user is strategic, how they interact with their content (e.g., probability of ``liking'' posts) does not simply depend on the piece of content itself; it also depends on the algorithm that recommended the content.
Thus, user strategization would have important managerial implications. 
Since recommendation algorithms are continually trained on user data, 
strategization can hurt platforms because users' revealed preferences do not necessarily reflect how they would behave under a different algorithm (or, more broadly, different circumstances).
Therefore, unless platforms take user strategization into account, 
they can be misled,
potentially to the detriment of both platforms and users.

\begin{figure}
    \centering
     \includegraphics[width=0.5\textwidth]{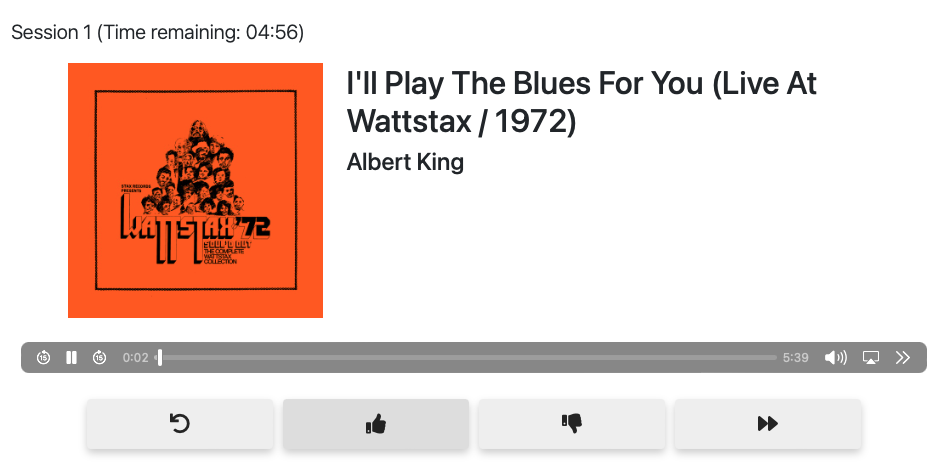}
 \caption{The music player interface with which participants interact.}
 \label{fig:player}
\end{figure}

\heading{Contributions and roadmap.}
Although strategization would have significant impacts on the data that platforms gather, to our knowledge, 
there has not been an experimental evidence investigating whether strategization in recommendation does indeed occur. {One potential reason for this lack of empirical work is 
that strategic behavior can be challenging for platforms to detect.
For example, in a theoretical model 
of strategization,
platform engagement is affected by both users' assessment of  recommended content \textit{and} by their understanding of the algorithm generating that content \citep{cen2023trust}. 
Platforms---whose recommendation algorithms are continuously trained on past user data---have no easy way of disentangling these two motivations for engagement, and thus no 
way to tease out the effect of strategization.

Our work threads this needle by testing for the presence of strategization in a lab experiment that uses a custom music player (shown in Figure \ref{fig:player}). In doing so, we were able to manipulate users' perceptions of the algorithm \textit{while controlling for the actual recommendations of the algorithm}, allowing us to empirically test for the presence of strategization for the first time (to our knowledge). This unique design allowed us to mimic the experience of interacting with a recommendation algorithm while avoiding the feedback loops that would cause confounding on real-world platforms.}

Using our custom music player, we 
ran a randomized control trial across 750 participants
that allowed us to observe how participants interacted with their songs (e.g., which songs participants ``liked'' and how long they listened to each song). %
Half of the participants were told that they would receive \textit{personalized} recommendations at the end of the study (see Figure \ref{fig:high_incentive_study_description}), and the other half were told that their behavior would be used to learn people's music preferences, but \emph{not} that they would receive recommendations (see Figure \ref{fig:low_incentive_study_description}).
Each participant then underwent two listening sessions. 
During the first session (the Warm-up), participants were asked to behave as they would on their typical music recommendation platform (e.g., Spotify). 
During the second, participants were 
randomly exposed to one of three descriptions of how our algorithm learns the participants' preferences (see Figure \ref{fig:info_conditions}).
This setup allowed us to test for user strategization along two axes: 
(i) whether participants believe their behaviors influence future recommendations, which we refer to as the Incentive condition; 
and 
(ii) the participants' beliefs about how their preferences are learned, 
which we refer to as the Information condition. 
That is, 
we sought to test whether participants who believe their actions affect their personalized recommendations adapt their behavior based on their knowledge of the algorithm's inner workings, 
i.e., whether they \emph{strategize}. 

Our findings show strong evidence of strategization. 
We first find that participants in different \emph{Information} conditions exhibit substantially different behaviors,
in terms of the number ``likes'' and ``dislikes'' that they use as well as the variance of time that participants ``dwell'' on each song. 
That is, 
participants adapt their behaviors based on their \emph{perception} of the algorithm's inner workings. 
Our second main finding is that participants in different \emph{Incentive} conditions also exhibit substantially different behaviors across almost all outcome metrics, 
which suggests that participants who believe that their behaviors influence future recommendations take actions (presumably) to improve them.
Together, these results provide strong evidence of strategization
and, notably, 
we observe significant differences in behavior even though the participant instructions provide only minor nudges to the participants' understanding of the system. 
Importantly, we discuss how our findings cannot be explained by experimenter demand effects 
(i.e., it is not simply the case that participants adapted their behavior to what they believed was desirable).

These effects are not just concentrated among highly active participants, but are observed across our outcome distributions, suggesting strategization is not a rare behavior. Furthermore, while we find that the nature and degree of strategization differ by individual characteristics, we observe consistent evidence of strategization even among participants one might expect to be naive (e.g., older participants). 
We further survey participants at the end of the study to understand whether and how users strategize on their own platforms,
and nearly half report definitive strategization.
These responses indicate that many users are highly aware of their recommendation algorithms. 
While a sizeable portion do not (consciously) strategize---stating that they do not wish to manipulate their algorithm---many others admit to ``gaming'' their algorithm. 
Several users report performing actions in ``Incognito'' mode, maintaining multiple user accounts for different ``moods,''
and obfuscating their preferences for many reasons, 
including not being ``pigeonholed'' by the algorithm.

\medskip 

The rest of the paper is organized as follows. Section \ref{sec:related_work} discusses the related work. Section \ref{sec:hypotheses} presents a formal definition for strategization and, in accordance, the hypotheses that we wish to test. Section \ref{sec:methods} describes the methodology and analysis we employ to test these hypotheses. Section \ref{sec:results} presents our results and evidence of strategization from the experiment and survey. Finally, we discuss the implications of our findings and future directions in Section \ref{sec:discussion}.

\subsection{Related Work}
\label{sec:related_work}

\heading{User awareness of recommendation algorithms.} Existing work shows evidence that 
users are {\em aware} of recommendation algorithms and have beliefs about how they work \citep{Newman2018-hl,Sirlin2021-hx,DeVito2021-se, eslami2016first, DeVito2018-it,Taylor2022-ch}.
The existence of these beliefs is a precursor to the phenomenon that we study, that users behave strategically in response to these beliefs. 

\heading{Qualitative evidence of user strategization in recommendation.}
More recently, there has been growing evidence of users attempting to \textit{influence} their future recommendation, primarily relying on self-reported survey data \citep{devito2017ruin, Shin2020-ip, Lee2022-wv, Simpson2022-vg, Haupt2023-yv}.
There are tutorials aimed to teach the general population how to  
``train'' their social media algorithms---that is, 
explicitly changing their behavior to induce a better feed
\citep{wsj2021investigation, Narayanan_2022_train, Narayanan_2023_understanding}. %
To the best of our knowledge, our work provides the first large-scale behavioral study that quantifies the existence of strategization by measuring observed behavior change, rather than relying on self-reported data.

\heading{Theoretical models of user strategization in recommendation.} 
Recent theoretical work propose models for user strategization (in recommendation and other personalization contexts) and use these models to illustrate how strategization affects both platforms and users \citep{cen2023trust, Haupt2023-yv}. \citet{Haupt2023-yv} study how strategization in recommendation can cause users to accentuate differences from other user preference profiles, 
affecting minorities in particular. %
\citet{cen2023trust} study conditions (e.g., mis-specification and misaligned incentives) under which strategization can mislead platforms and, in turn, hurt users.
We adapt the model from the latter to develop testable hypotheses in Section \ref{sec:hypotheses} for the existence of strategization and analyze consumption patterns in addition to user feedback (``likes'' and ``dislikes''). Our work provides empirical evidence for ``long-term planning'' as a mechanism leading to strategization.

\heading{Strategization in other contexts.}
We note that in the context of recommendation, there is an existing body of work showing that \textit{content creators} strategize in terms of the type and frequency of the content they create \citep{arriagada2020you, huang2022causal, mummalaneni2023content, immorlica2024clickbait, jagadeesan2022supply, hron2023modeling, huttenlocher2023matching}.
We consider our focus of user-side strategization as a distinct phenomenon from creator-side strategization, as well as documented strategization in online 
auctions \citep{Edelman2007-sx} and freelancing \citep{Rahman2021-bc}, and
ride-share platforms \citep{Marshall2020-rn}.
This user-side strategization distinctly aims to assist the algorithm in learning the user's own preferences. 

We further distinguish our work from strategic classification \citep{bruckner2009nash, Hardt2016-ks} and {generalized} strategic classification introduced in \citet{levanon2022generalized}. 
In the original strategic classification formulation, agents strategize to induce a positive decision (e.g., loan approval) whereas agents in the generalized strategic classification model strategize to induce the \textit{correct} decision. 
Our formulation is closed to the latter in the sense that recommender systems seek to personalize to user preferences. 
However, our work is distinct in that we examine settings with repeated interactions, where an individual is cognizant of the fact that their current actions are used as training data and may therefore influence future outcomes, 
rather than a one shot prediction setting.

\heading{Learning from revealed preferences.} Our work contributes to the observation that users' revealed preferences (observed behavior) may not be indicative of their true interests \citep{beshears2008preferences}. In the recommender system space, \citet{Kleinberg2022-wy, morewedge2023human} suggest that relying on revealed preferences can result in suboptimal recommendations, due to factors like users' inconsistent preferences and habitual behavior. Our work shows how revealed preferences depend not only on users' true interests, but also their beliefs about the algorithm and whether they are forward looking.

\section{Model and Hypotheses}\label{sec:hypotheses}

In this section, we outline the precise hypotheses that guide 
the design of our experiment.
Recall that our goal is to test for the presence (or lack) of 
{\em user strategization} in recommendation.
To that end, 
we first describe an intuitive notion of strategization.
In Section \ref{ssec:model}, we present a model that aims to capture this notion. We then use this model to formulate two testable hypotheses (Section \ref{ssec:hypotheses}) that we then test in an online lab experiment (Section \ref{sec:methods}). 

The intuitive notion of strategization that we focus on here is what one 
might call \textit{forward-looking} \citep{cen2023trust}.
Under the definition we adopt,  
a strategic user adapts their behaviors 
based on their belief about how the recommendation algorithm uses these behaviors to generate future recommendations. 
For example, consider a user browsing Instagram and deciding 
whether to click on a specific video. 
The user may be interested in watching the video, 
but at the same time might believe that clicking on the video  
will cause the algorithm to over-recommend similar content. 
The user may {\em strategically} refrain from clicking on the video---in contrast, a {\em naive} user who does not take their future recommendations into account may still click on the video. 
In this way, strategic users select their current actions 
in an effort to elicit better downstream recommendations.\footnote{Note that this phenomenon is distinct from the inconsistent preferences (e.g., ``junk'' versus ``healthy'' content preferences) phenomenon highlighted by \citet{Kleinberg2022-wy}. In this work, we consider strategization with respect a user's understanding of the algorithm and how they can communicate with the algorithm, whereas \citet{Kleinberg2022-wy} consider behavior induced by a user's own internal preference inconsistencies. %
}%

{There are two elements that are key to this definition of strategization. 
The first is that the user is aware of the algorithm and develops a belief about how the algorithm works. 
The second is that the user adapts their behavior according to this belief in order to receive better downstream recommendations.
These two elements form the basis for the two hypotheses that we test in our lab experiment. In the following, we describe our definition and the resulting hypotheses in detail.}

\subsection{Model of Strategic Users}
\label{ssec:model}
We adapt the model of \citet{cen2023trust},
distilled to the components that we test experimentally in this work
and adjusted for a finite-horizon of interactions with the recommendation algorithm.

\heading{The platform.}
Let $\cZ$ be the set of content available on a platform,
and let $\mathcal{B}$ be the set of ways (or behaviors)
with which a user can respond to a piece of content $Z \in \mathcal{Z}$.
In the setting of music recommendation, 
$\cZ$ is the set of songs the platform can recommend, 
and
a behavior $B \in \mathcal{B}$ could be the number of seconds that a user listens to a song. 
A behavior could also correspond to explicit feedback (e.g., in the form of ``likes'' or ``dislikes''),
or it could correspond to a combination of various actions (e.g., ``listen for 5 seconds and click the like button'' could be one behavior and ``listen for 50 seconds and click the dislike button'' could be another).

The user and platform engage in $T > 0$ repeated interactions where, 
at each 
time step $t \in \{1, \ldots, T\}$,
the platform gives a {\em recommendation} $Z_t \in \cZ$ 
and the user responds with a behavior $B_t \in \mathcal{B}$.
Based on both the recommendation $Z_t$ and their behavior $B_t$, 
the user collects a payoff 
$U(Z_t, B_t)$. For example, given a song recommendation $Z_t$ that the user likes, we may have $$U(Z_t,  \text{ ``listen for 5 seconds and click like''}) < U(Z_t, \text{ ``listen for 50 seconds and click like''}).$$ 
We note that the user may receive the same payoff for different actions, e.g., $U(Z_t, \text{ ``listen for 50 seconds and click like''}) = U(Z_t, \text{ ``listen for 50 seconds and click dislike''})$, 
since clicking ``like'' or ``dislike'' may not affect the user's enjoyment when listening to the song, 
though it may affect the algorithm's ability to personalize future song recommendations, 
as we discuss next.

\heading{The (user's belief about the)  recommendation algorithm.}
The user believes that the platform generates its recommendations 
using an {\em algorithm} $\mathcal{A}$,
which maps the interaction history 
$\mathcal{H}_t = \{(Z_1, B_1), \ldots, (Z_{t}, B_{t})\}$ at time $t$
to a new recommendation $Z_{t+1}$ at time $t+1$. 
Based on the algorithm $\mathcal{A}$ and 
the system parameters $\mathcal{B}, \mathcal{Z},$ and $T$,
the user chooses a {\em behavior policy}
$\pi(\cdot \,; \mathcal{A}, \mathcal{B}, \mathcal{Z}, T)$ 
that maps a recommendation $Z_t$ 
to a distribution over behavior.\footnote{We consider users that are 
{\em stationary} in that they do 
not use the history $H_t$. This is 
without loss of generality, given that 
we do not restrict $\mathcal{Z}$, 
which can be adjusted to ``contain'' the history.} 
At time $t$, the user responds to 
the recommendation 
$Z_t$ by sampling $B_t \sim \pi(Z_t; \mathcal{A}, \mathcal{B}, \mathcal{Z}, T)$.

Note that for the purposes of this model, the way the platform {\em actually}
generates recommendations is irrelevant, and in particular, 
it need not match the user's belief.
In the subsequent discussion, we consider a setting where a user's belief about how the algorithm works is fixed across time. It is an interesting question for future work to study how the user learns this belief.

\heading{The naive policy.}
The simplest behavior policy is the {\em naive} policy, which responds 
to recommendation $Z_t$ with the behavior that maximizes the user's immediate
utility, i.e.,
\begin{equation}
    \label{eq:naive_policy}
    \pi^{\text{naive}}(Z_t;  \mathcal{B}) 
    = \delta\left\{\arg\max_{B \in \cB}\ U(B, Z_t)\right\},
\end{equation}
where $\delta\{\cdot\}$ is the Dirac delta distribution 
(i.e., a degenerate distribution which places probability one on 
its argument and probability zero on everything else). 
In the language of recommender systems, 
the naive policy corresponds to a user's ``ground-truth'' content preferences;
A platform typically assumes that the user follows the naive policy and refers to it as the user's ``ground-truth'' behavior.
In particular, note that $\pi^{\text{naive}}$ is \emph{exogenous} in that it
does not depend on
either the  perceived algorithm $\mathcal{A}$, nor the time horizon $T$. 

\heading{The strategic policy.}
However, a user may indeed account for $\mathcal{A}$ and $T$ when deciding how to behave. 
They may, for example, choose sub-optimal short-term behavior to elicit better recommendations from $\mathcal{A}$ in the long term. 
Such a policy depends on both $\mathcal{A}$ and $T$. 
To capture this phenomenon, 
we define the following class of \emph{strategic policies} where the user chooses a strategy that optimizes their utility
over $T$ time steps for $T > 1$:
\begin{equation}
    \label{eq:strategic_policy}
    \pi^{\text{strat}}(\cdot;\mathcal{A}, \mathcal{B}, \mathcal{Z}, T) = \arg\max_{\pi}\ \mathbb{E}_{\mathcal{H}_T(\pi, \mathcal{A})} 
                    \left[\sum_{t=1}^T U(B_t, Z_t)\right],
\end{equation}
where $\mathcal{H}_T(\pi, \mathcal{A})$ is a function mapping from a user policy and a 
platform algorithm to the (endogenous) $T$-step rollout of recommendations and actions 
generated by the user and platform interacting 
according to $\pi$ and $\mathcal{A}$
respectively.  
For simplicity, 
we assume that the user chooses their strategic policy once;
given knowledge of $\mathcal{A}$ and $T$, 
the strategic user computes $\pi^{\text{strat}}(\cdot;\mathcal{A}, \mathcal{B}, \mathcal{Z}, T)$ before $t = 1$. 
Note that this policy class reduces to the naive policy \eqref{eq:naive_policy} for $T=1$.%

\heading{Example.} %
Consider the Instagram example given at the start of this section, 
in which a user is interested in a video but does not want to indicate too much preference for it in order to avoid being over-recommended similar content down the line.
Suppose, for instance, that a user is recommended a cat video at time $t=1$, where they would receive positive payoff for engaging %
with the video.
Suppose, however, that the user believes indicating interest in the video will 
cause the platform to show only cat videos at times $t=2, \ldots, T$, and the user prefers many other types of videos over cat videos.
The user may believe they have lower expected utility at times $t=2, \ldots, T$ if they engage with the video now and thus prefer to not engage at $t=1$ so that they receive better recommendations in the future. 
Such a user would be \emph{strategic} under the definition above in that they choose a policy that optimizes utility over a time horizon $T > 1$.

A strategic user's behavior also depends on their belief about how the recommendation algorithm selects future content. 
For example, a strategic user who believes that Instagram tracks time spent engaging with a video (or ``dwell time'') might watch the cat video for a shorter amount of time than they would in the absence of an algorithm. 
In contrast, a strategic user who believes that Instagram only tracks explicit interactions (e.g., ``likes'') but not dwell time may watch the cat video but refrain from liking or commenting. 
In this way, two users with the same utility for a given video may behave differently due to different understandings of the algorithm's inner workings \emph{if they are strategic}. 
Our definition above also accounts for the fact that strategic users' behaviors depend on their perceptions of the algorithm.\footnote{Note that we consider strategization with respect to a utility function defined over $(Z,B)$ pairs, which does not capture the setting where users prefer a diverse set of recommendations. A natural extension is to consider a higher dimensional utility function that depends on the sequence of content shown.}

\medskip

\subsection{Testable Hypotheses}
\label{ssec:hypotheses}

Testing for user strategization is challenging. 
For one, 
each user's true preferences $U$ are unknown, 
making it difficult to determine whether their revealed preferences (i.e., what is observable) matches their true preferences.
For another, users have heterogeneous preferences (i.e., there is a different, hidden $U$ for each user). 
Moreover, a strategic user's behavior depends
on their mental model of their algorithm,
which can vary wildly from user to user.

Despite these challenges, 
the model described above suggests two key differences between the behaviors of the naive user and the strategic user, which we formalize into testable hypotheses.
We make two observations:
\begin{enumerate}
    \item The naive policy does not depend on (the user's belief of)
    the platform's recommendation algorithm $\mathcal{A}$, while the strategic 
    policy changes for different beliefs $\mathcal{A}$.
    \item The naive policy is independent of the time horizon $T$ of the 
    platform-user interaction, whereas the strategic policy is not.
\end{enumerate}
Thus evidence that the behaviors of users with different beliefs about the algorithm or different beliefs about the time horizon are different would, holding all else equal, suggest the presence of strategization.
Next, we translate these two observations into concrete hypotheses about user behavior, by varying the user's perceived algorithm 
$\mathcal{A}$ and the time horizon $T$ exogenously, and estimating the 
effect on $\pi^*(\cdot;\mathcal{A}, \mathcal{B}, \mathcal{Z}, T)$ from samples. 
\begin{hypothesis}[Information Condition]
\label{hyp:information_formal}
    Holding all else constant, 
    changing the participants' beliefs about the recommendation algorithm changes the way they behave. 
    Formally, there exist $\mathcal{A}$ and $\mathcal{A}'$ such that 
    $\pi^*(\cdot;\mathcal{A}, \mathcal{B}, \mathcal{Z}, T) \neq \pi^*(\cdot;\mathcal{A}', \mathcal{B}, \mathcal{Z}, T)$.
\end{hypothesis}

\begin{hypothesis}[Incentive Condition]
    \label{hyp:incentive_formal}
    Holding all else constant, changing the time horizon 
    of the platform-user interaction will change the participant's behavior.
    Formally, there exist time horizons $T_1$ and $T_2$ such that
    $\pi^*(\cdot;\mathcal{A}, \mathcal{B}, \mathcal{Z}, T_1) \neq \pi^*(\cdot;\mathcal{A}, \mathcal{B}, \mathcal{Z}, T_2).$
\end{hypothesis}
The first hypothesis implies that users are not only aware of their algorithm, 
but also adapt their behavior based on their understanding of the algorithm. 
The second hypothesis implies that users adapt their behavior \emph{because} they are aware that their current behavior influences future recommendations.
Together, 
these hypotheses would indicate that users are strategic in that they \emph{adapt} their behavior based on their understanding of the algorithm in order to elicit good \emph{future} payoffs. %
Our study---a controlled lab experiment---enables us to test the two hypotheses above  {\em directly} by influencing users' beliefs about how the algorithm works and the relevant time horizon.

Note that---while we use the model in \cref{ssec:model} to formalize the hypotheses and effects that we expect---we do not explicitly fit the model to observed user behavior 
(i.e., we do not fit $\pi$ or $U$).%

\section{Methodology}\label{sec:methods}

We built a custom music player that allows participants to listen to and interact with songs, similar to how they would on Spotify or Pandora (see Figure \ref{fig:player}). 
{The interface allows us to measure data on each participant's behavior (e.g., the number of times they click ``like'', ``skip'', and ``replay'' when presented with different songs). Importantly, we can also include subtle messaging changes that influence participants' beliefs about the algorithm and about the time horizon. 

Using this interface, we conducted a behavioral experiment to 
test the hypotheses formalized in Section \ref{sec:hypotheses}.} 
Specifically, recall that our goal is to answer the questions: 
Do different Information and Incentive conditions affect participant behavior in a systematic way? If so, do the observations support the strategization hypotheses given in Section \ref{sec:hypotheses}?
We additionally asked participants to complete a post-experiment survey on whether they intentionally strategize on real-world platforms such as 
Spotify and TikTok.

All analyses are pre-registered, except where they are designated ``post-hoc.'' Our pre-registration and analysis plan is available at \url{https://aspredicted.org/WVF_6SH}.

\subsection{Participants}
We recruited 750 participants from 
CloudResearch Connect.
Of the recruited
participants, we excluded 
28
participants who ran into 
technical issues.
Of the remaining participants, 
50
failed at least two 
audio-visual attention checks or written attention 
checks. 
Finally, another 15 participants had metrics (likes, 
dislikes, skips, and dwell time) that were more than four standard deviations away from the average.
The final sample has 
657
participants, 
of which 48\% are male, 52\% are female; 
45\% are 35 years-old and below, and 70\% use
Spotify or music recommendations platforms at least a few times a week. 
In accordance with the standards of MIT's Institutional Review Board (IRB), this study was granted an exemption from full IRB review on March 9, 2023.

\subsection{Music Platform}

\begin{figure}
     \centering
     \hfill
     \begin{subfigure}[b]{0.49\textwidth}
         \centering
         \includegraphics[width=\textwidth]{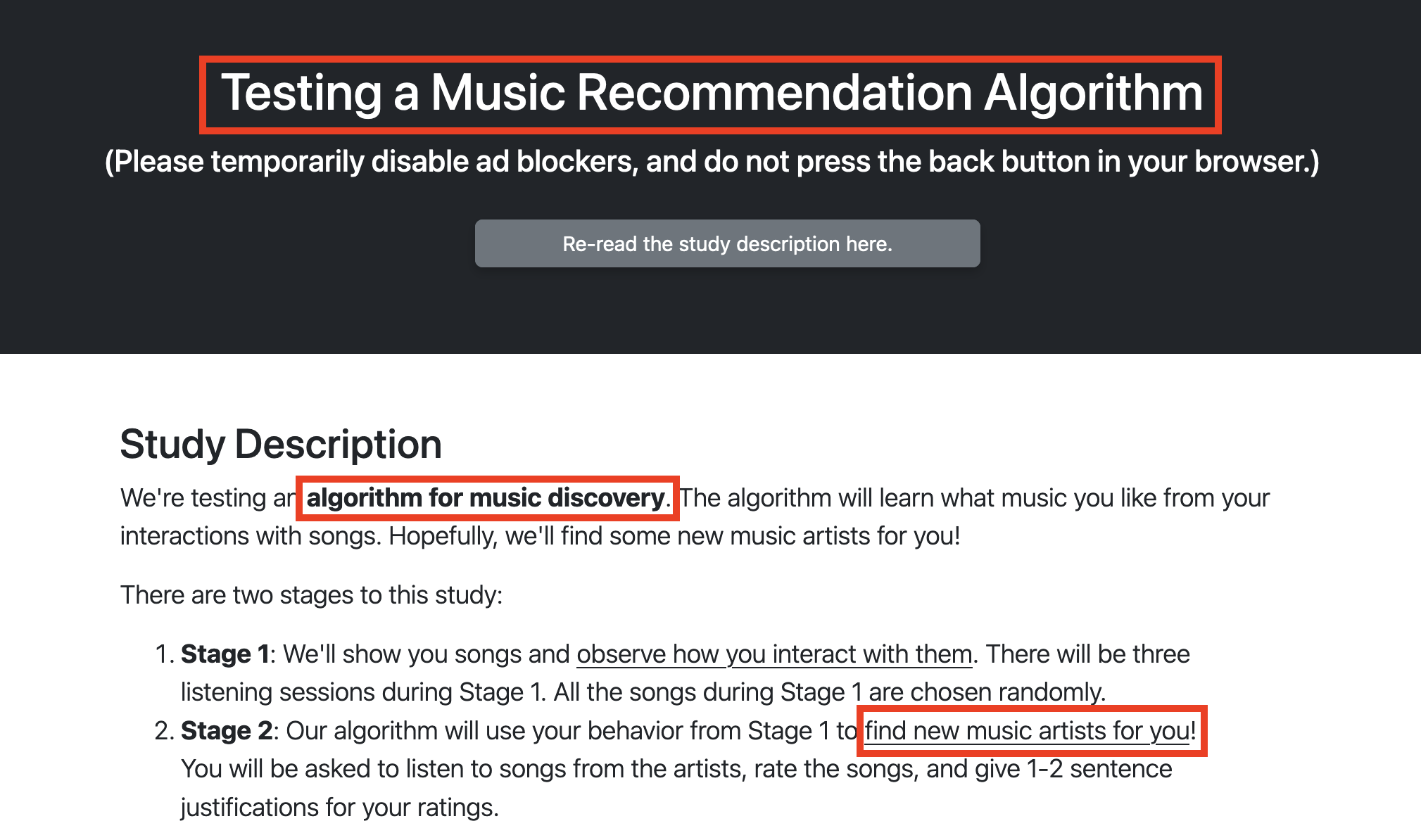}
         \caption{``Treatment'' Incentive study description}
         \label{fig:high_incentive_study_description}
     \end{subfigure}
     \hfill
     \begin{subfigure}[b]{0.49\textwidth}
         \centering
         \includegraphics[width=\textwidth]{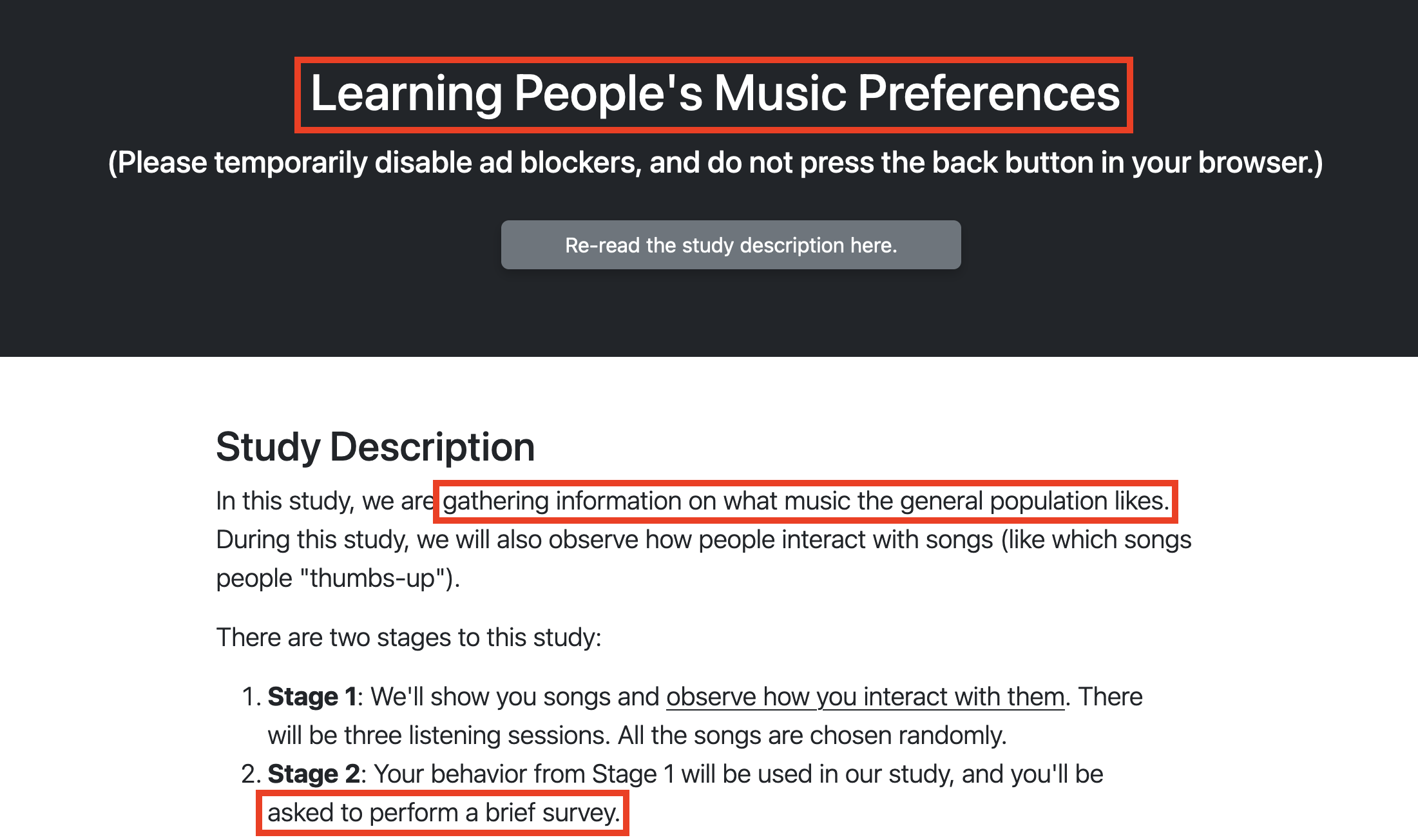}
         \caption{``Control'' Incentive study condition}
         \label{fig:low_incentive_study_description}
     \end{subfigure}
     \vspace{12pt}
        \caption{%
        (a) The study description that participants in the ``Treatment'' Incentive condition see. These participants are told that their behaviors will be used to generate personalized music recommendations at the end of the study. (b) The study description that participants in the ``Control'' Incentive condition see. These participants are told that their behaviors are used to learn what music the general population likes. The participants are randomly divided into the ``Treatment''and ``Control'' Incentive conditions.}
        \label{fig:setup_player_info_ex}
\end{figure}

\begin{figure}
     \centering
     \begin{subfigure}[b]{0.46\textwidth}
         \centering
         \includegraphics[width=\textwidth]{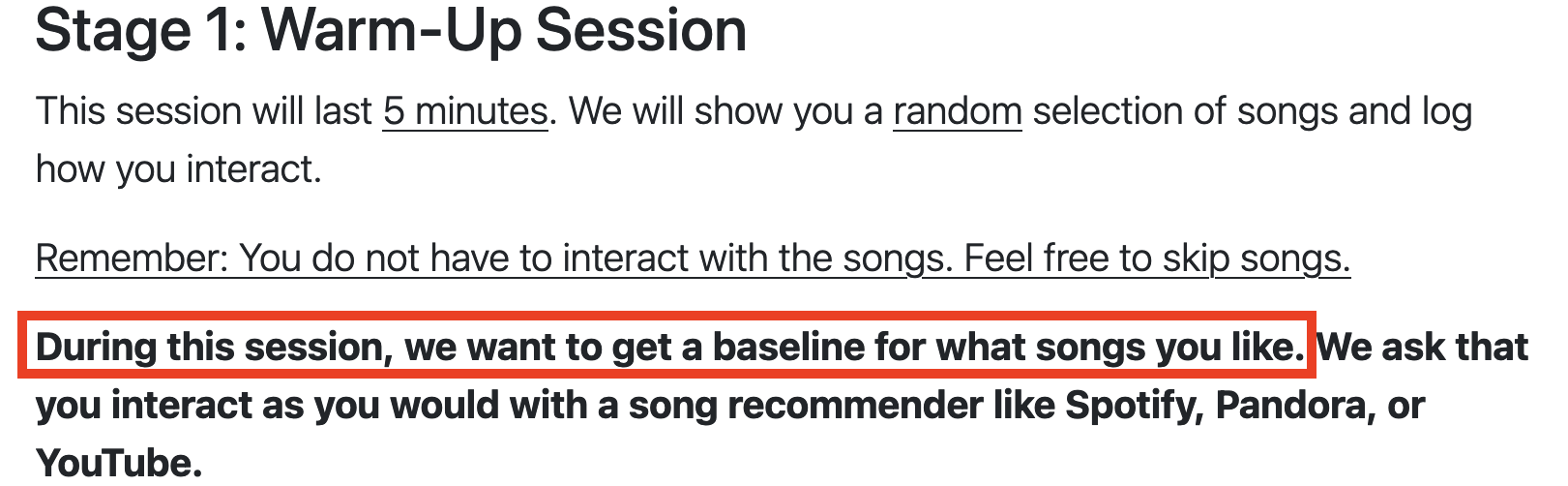}
         \caption{For Warm-up session}
         \label{fig:info_condition_Warm-up}
     \end{subfigure}
     \hfill
     \begin{subfigure}[b]{0.46\textwidth}
         \centering
         \includegraphics[width=\textwidth]{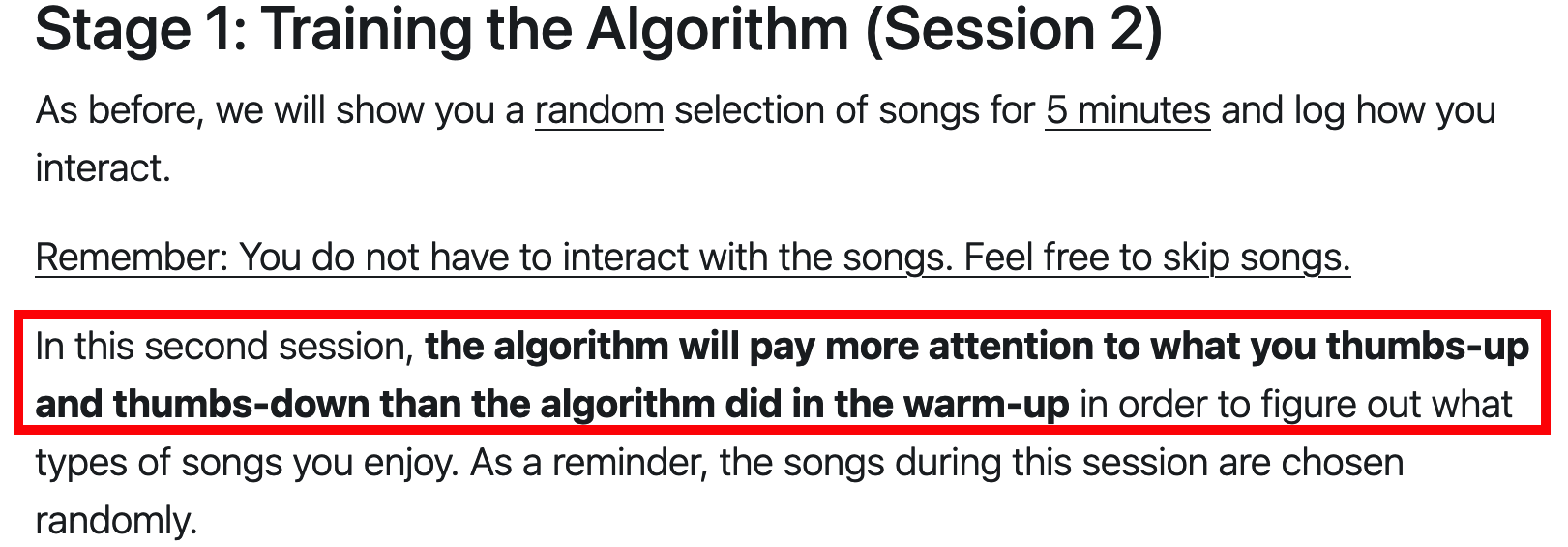}
         \caption{For Session 1 (``Likes'' condition)}
         \label{fig:info_condition_likes}
     \end{subfigure}
     \\
     \vspace{12pt}
     \begin{subfigure}[b]{0.46\textwidth}
         \centering
         \includegraphics[width=\textwidth]{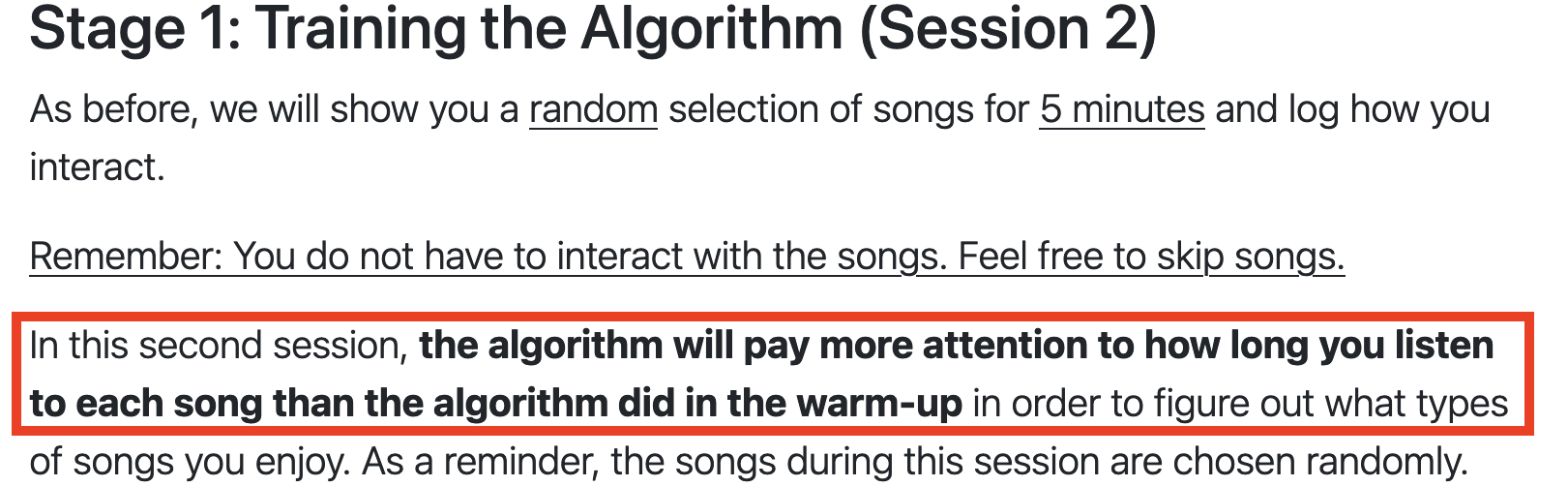}
         \caption{For Session 1 (``Dwell'' condition)}
         \label{fig:info_condition_dwell}
     \end{subfigure}
     \hfill
     \begin{subfigure}[b]{0.46\textwidth}
         \centering
         \includegraphics[width=\textwidth]{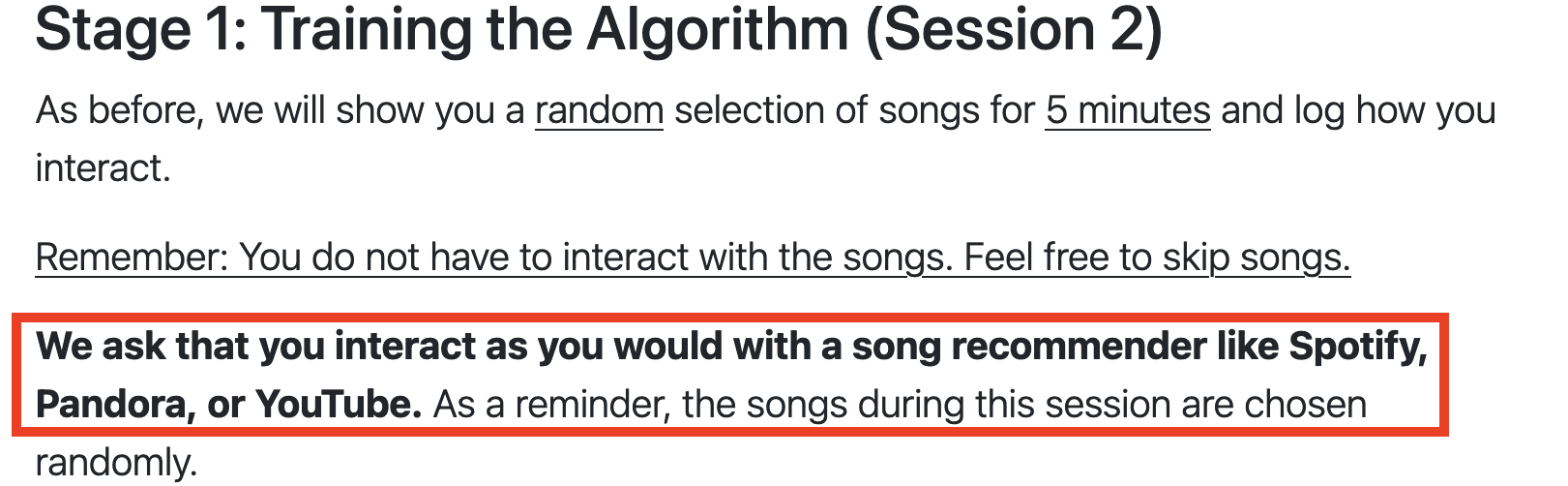}
         \caption{For Session 1 (``Control'' condition)}
         \label{fig:info_condition_neutral}
     \end{subfigure}
     \vspace{4pt}
        \caption{Participants undergo two listening sessions. The first session for all participants is the Warm-up session, as shown in (a). 
        In the second session, 
        participants are randomly assigned to one of three Information conditions, 
        as shown in (b), (c), and (d). 
        (The descriptions above are shown to participants in the ``Treatment'' Incentive condition. The ``Control'' Incentive descriptions are analogous.)}
        \label{fig:info_conditions}
\end{figure}

We built a custom music player on which participants can listen to and interact with songs. 
Each participant underwent two 5-minute listening sessions. 
During each session, 
participants could ``like,'' 
``dislike,'' 
``skip,'' and ``restart'' each song
as well as skip ahead to any time within a song, 
as shown in \cref{fig:player}.
We logged all participants' actions that involve clicks (such as the actions listed above). 

We constructed a song bank of 196 songs purchased from iTunes on March 5, 2023 from the ``Top Songs'' of 16 genres. 
We partitioned the songs into three groups---one for each listening session, and a third for an additional listening session collected for secondary analysis. In each listening session,
we presented participants with were chosen uniformly at random (without replacement) from the corresponding group. 
Note that the presented songs thus did \emph{not} depend on the participants' interaction behavior,
as described next.%

\subsection{Experimental Conditions}\label{sec:conditions}
We used a 3-by-2 factorial design with the conditions described below.
The first factor, 
which we refer to as the Information condition, 
tests \cref{hyp:information_formal}.
The second factor, 
which we refer to as the Incentive condition,
tests \cref{hyp:incentive_formal}.
In total, there were thus six different pairs of conditions that influence each participant's knowledge about (i) the algorithm used to learn their preferences and (ii) whether the learned preferences are used to generate personalized recommendations for the participant. 
Since participants are first assigned to an Incentive condition, then an Information condition, 
we describe the conditions in that order. 

\heading{Study description.}
At the start of the study, 
participants were randomly provided one of two descriptions of the study's purpose. Each description corresponded to an 
\emph{Incentive condition} (see Figure \ref{fig:setup_player_info_ex}). 
In particular, each participant was assigned one of two possible Incentive conditions:
\begin{itemize}
    \item {\bf ``Control'' Incentive condition.} Participants in this  condition were told that the goal of 
    the study was to learn what music the general population likes based on their interactions with songs. 
    They were told that they would first undergo listening sessions, then be asked to answer a brief survey. 

    \item {\bf ``Treatment'' Incentive condition.}
    Participants in this condition were told that the goal of the study was to test an algorithm for music discovery that will try to learn their preferences from their interactions with songs. 
    They were told that they would first undergo listening sessions, then be asked to give feedback on songs that our algorithm recommends. 
\end{itemize}
The Incentive conditions are intended to test whether participants plan ahead when they know that their current actions affect their future recommendations, as per \cref{hyp:incentive_formal}.\footnote{In our experiment, songs were generated randomly (not personalized) until the very end (and even then, only if the participant is in the ``Treatment'' Incentive condition. Therefore, under our formalization, one can view $\mathcal{A}$ as random until $t = T$.} {Compared to a participant in the ``Control'' Incentive condition who undergoes two listening sessions, a participant in the ``Treatment'' Incentive condition who undergoes the same two listening sessions and \textit{afterward} receives personalized recommendations (``new music artists for you!'') effectively has a longer time horizon and may alter their behavior in the listening sessions to ensure good personalized recommendations at the end.}

Note that participants in the ``Control'' Incentive condition may plan ahead to some degree if they naturally strategize in the wild, because planning ahead on recommendation platforms has become habitual for them.
If this does occur, it would dampen our observed treatment effect. 
If, however, we still observe a treatment effect despite this spillover, 
it would provide even stronger evidence in support of \cref{hyp:incentive_formal}.

\heading{First session.}
The first listening session was the ``Warm-up session.'' 
For this session, 
\emph{all} participants were told to interact as they would with a song recommender like Spotify or Pandora in order for us to get a baseline for the music they like. 

\heading{Second session.}
Independently of their assigned Incentive condition, each participant was
randomly assigned one of three {\em Information conditions}.
Just before the start of the second session, each participant received 
some information dictated by their  Information condition
(See Figure \ref{fig:info_conditions}):
\begin{itemize}
    \item {\bf ``Control'' Information condition.} As in the Warm-up, participants in this condition received no information about how their preferences are learned. They were told to interact as they would with Spotify or Pandora. 
    
    \item {\bf ``Likes'' Information condition.} 
    Participants in this condition were told that, in order to learn their music preferences, the algorithm would pay more attention to how they ``like'' (thumbs-up) and ``dislike'' (thumbs-down) songs as compared to the Warm-up session. 
    
    \item {\bf ``Dwell'' Information.} Participants in this condition were told that, in order to learn their music preferences, the algorithm would pay more attention to their dwell time (how much time they spend on each song) as compared to the Warm-up session. 
\end{itemize}
To summarize, 
some participants underwent the ``Control'' Information condition for both listening sessions, 
some underwent the ``Control'' Information  \emph{then} the ``Likes'' Information condition, 
and the rest underwent the ``Control'' Information  \emph{then} the ``Dwell'' Information condition. 

Intuitively, we expect a participant that is strategic in the sense of 
Hypothesis \ref{hyp:information_formal} to 
convey their preferences to the algorithm in a way that depends on their 
Information condition. 
For example, a strategic participant in the ``Likes'' Information condition
would more frequently use ``likes'' and ``dislikes.'' Similarly,
we would expect a strategic participant in the ``Dwell'' Information condition to convey their preferences by manipulating their dwell time, either by listening to songs that want recommended for longer or more quickly skipping songs that they do not want recommended. 

\medskip

Note that the way we generated songs did \emph{not} change across participants (all songs during the listening sessions were generated randomly from the same pool of songs). 
We only change the {\em instructions} that participants receive. {This is necessary for clean identification of the effects of the Information and Incentive conditions; if songs in the listening sessions depended on past participant behavior, participants' behaviors would be confounded with a changing sequence of songs.}

\subsection{Post-Experiment Survey}\label{sec:post-experiment-survey-methods}

At the end of the study, 
all participants were asked to complete a survey.
The full list of questions is given in \cref{app:survey_Qs}.
In addition to demographic information,
we asked participants several multiple-choice/checkbox questions to query: (1) whether they changed the way they interacted across sessions and, if so, how; 
(2) how they believe their recommendation algorithms work on Spotify, Facebook, etc.; 
and 
(3) how much time they spend online. 
In addition, 
we ask one open-ended text question:
\emph{Do you ever try to “talk” to your algorithm or “hide” things from it? For example, do you ever give a song a “thumbs-up” just to tell Spotify that you want to see similar songs? Or do you sometimes avoid clicking on an advertisement just because you’re worried about getting many similar advertisements in the future? If you do, tell us how and why.}

\subsection{Analysis}
We examine the data collected from our experimental procedure for 
signs of strategization.
To test 
Hypotheses \ref{hyp:information_formal}
and \ref{hyp:incentive_formal} 
from \cref{sec:hypotheses},
we look at {\em average treatment effects} of the Information conditions 
and Incentive conditions on participant behavior. We then look at how these effects manifest across our outcome distributions, and whether there are heterogeneous effects by individual-level characteristics. Finally, we analyze the post-experiment survey that participants took in order to get a qualitative picture of strategization.

\subsubsection{Outcome variables.}\label{sec:outcome_vars}

We pre-registered the following outcome variables.

\begin{enumerate}
    \item \textbf{Likes + Dislikes}. The number of songs that the participant has either ``liked'' (thumbs-up) or ``disliked'' (thumbs-down) during the session.
    \item \textbf{Fast Skips}. The number of times that the participant ``skips'' a song during the first 5 seconds of the song during the session.
    \item \textbf{Dashboard Clicks}. The number of times that the participant clicks on the song player dashboard. Includes clicks on the ``back'' and ``pause'' buttons, 
    as well as ``skipahead'' events (where the participants scroll forward or backward through the song). We exclude clicks on the ``like,'' ``dislike,'' and ``skip'' buttons, since these actions are captured by our other outcomes variables. 
    \item \textbf{Log Average Dwell Time}. Average length of time participant listens to each recommended song in milliseconds, log-scaled.
    \item \textbf{Log Standard of Deviation Dwell Time}. Standard deviation of the time participant listens to each recommended song in milliseconds, log-scaled.
\end{enumerate}
Additionally, we pre-registered an analysis examining the {\em proportion} (rather than the count) of (i) ``likes'' and ``dislikes'' and (ii) ``fast skips'' per song listened. Due to space constraints, we report these results in Appendix \ref{app:additional_experimental_details}; they do not change the interpretation of our findings. 

\subsubsection{Group means and average treatment effects.}\label{sec:methods-ate}

To test for the presence of strategization, we examine how the outcome variables (averaged across participants) in the test listening session differ across our (i) Information and (ii) Incentive conditions. 

For each of our outcome variables, we fit a model with the respective outcome variable of interest as our dependent variable and treatment dummies for (i) the Incentive condition $D_{\text{Incentive}}$, (ii) the Information condition $D_{\text{Information}}$, and (iii) their interaction. 
In addition, 
we fit an additional specification that includes participants' pre-treatment behavior $X_{\text{pre}}$ in the Warm-up session as a control variable (e.g., we include the number of ``likes'' and ``dislikes'' in the Warm-up Session as a control when Likes + Dislikes is our outcome of interest). 
In other words, 
we fit the outcome variable (with the appropriate model, as specified next) to the following:
\begin{align*}
    \beta_0 + \beta_1 D_{\text{Incentive}} + \beta_2 D_{\text{Information}} + \beta_3 (D_{\text{Incentive}} \times D_{\text{Information}}) + \beta_4 X_{\text{pre}} + \varepsilon .
\end{align*}
We now specify the models used for each of the outcome variables in Section \ref{sec:outcome_vars}.
For our three count variables, (i) Likes+Dislikes, (ii) Fast Skips, and (iii) Dashboard Clicks, we use a Poisson quasi-maximum-likelihood (quasi-Poisson) regression. We use a quasi-maximum-likelihood model in order to account for potential overdispersion in the engagement data \citep{wooldridge_quasi-likelihood_1999}. For our continuous dwell time variables, (i) Log Average Dwell Time and (ii) Log Standard Deviation of Dwell Time, we use an OLS regression. 

To report interpretable versions of the main effects of each condition, we calculate the average marginal effect (AME) of each treatment condition compared to the respective control group in that condition. For example, we report the difference in Likes + Dislikes predicted by our model between the ``Treatment'' Incentive condition and the ``Control'' Incentive condition, pooling across all levels of the Information condition.
We control the false discovery rate across multiple comparisons using the Benjamini-Hochberg procedure.

\subsubsection{Subgroup means and treatment effects.}
In our pre-registration, we also stated we would examine potential heterogeneous effects among (i) participants younger than 25 years old and (ii) those who use TikTok, since we hypothesized that these subgroups might be more prone to strategic behavior. However, because we only had 72 participants below 25 (as our participants must also be at least 18 years old), we instead chose to look at participants who were below 35 years old for greater statistical power. 
Also, due to an error our final survey did not include a question specifically about TikTok (although we do ask about online platform use). 
We therefore divide our participants based on a different question where we ask participants about their use of music recommendation platforms, 
as this question closely aligns with our experimental setup.
We refer to this question as Spotify Use for brevity, and 
we code Spotify Use greater than once per week as ``Often,'' and less than or equal to once per week as ``Rare.''

We then calculate the conditional average treatment effect (CATE) for each subgroup of interest, using the same methodology described in Section~\ref{sec:methods-ate}. We test for heterogeneity in our treatment effects across subgroups by examining the difference-in-CATEs (DICs) using a Wald Test.

\section{Results}\label{sec:results}

In this section, 
we present and discuss our main findings.
We provide further results in Appendices \ref{app:additional_experimental_details} and \ref{app:additional_survey_results}.
We find strong evidence supporting both \cref{hyp:information_formal}
 and \cref{hyp:incentive_formal}. 
 We also find that Age and Spotify Use do not moderate the effects of the Information condition and mildly moderate the effects of the Incentive condition, suggesting that strategization occurs across subgroups but is mildly more prominent among participants who are expected to gain more from strategizing ``in the wild.''
 In \cref{sec:survey}, 
 we analyze the results of our post-experiment survey, focusing on free-form responses given by users on whether and why they strategize in recommendation.

\subsection{Do People Strategize?}\label{sec:exp_results}

\begin{figure}[t]
    \centering
    \includegraphics[width = 0.9\textwidth]{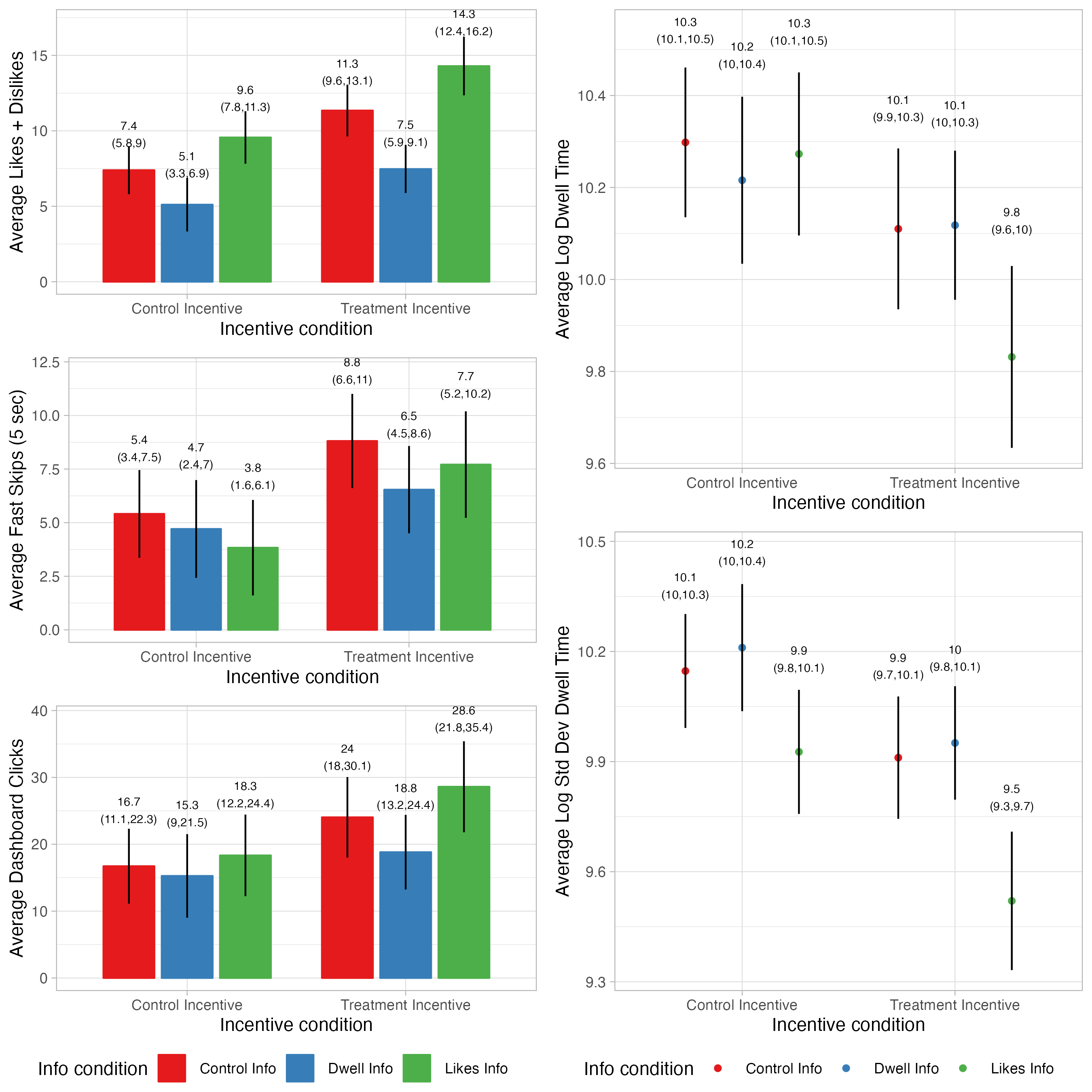}
        \caption{Means and 95\% confidence intervals (CIs) across our conditions for our five outcome variables of interest, 
        as described in Section \ref{sec:outcome_vars} and using the models in Section \ref{sec:methods-ate}.}
\label{fig:average-by-condition}
\end{figure}

Our results provide strong evidence that users strategize when interacting with recommendation algorithms.
We find that participants behave differently depending on (i) their belief about how the  algorithm learns their preferences (Hypothesis \ref{hyp:information_formal}) and (ii) whether they believe their behaviors affect future outcomes on the platform (Hypothesis \ref{hyp:incentive_formal}). 
These findings suggest not only that user behavior is algorithm-dependent, 
but also that this algorithm-dependence is driven by users' understanding that current actions affect downstream recommendations.
Figure~\ref{fig:average-by-condition} summarizes our main results, 
showing the means of our outcome variables across our Information and Incentive conditions, as described in Sections \ref{sec:outcome_vars} and \ref{sec:methods-ate}. 
We now discuss these results  in light of our two Hypotheses: \textbf{H1} (which corresponds to Hypothesis \ref{hyp:information_formal}) and \textbf{H2} (which corresponds to Hypothesis \ref{hyp:incentive_formal}). 

\begin{figure}[t]
    \centering
    \includegraphics[width = 0.9\textwidth]{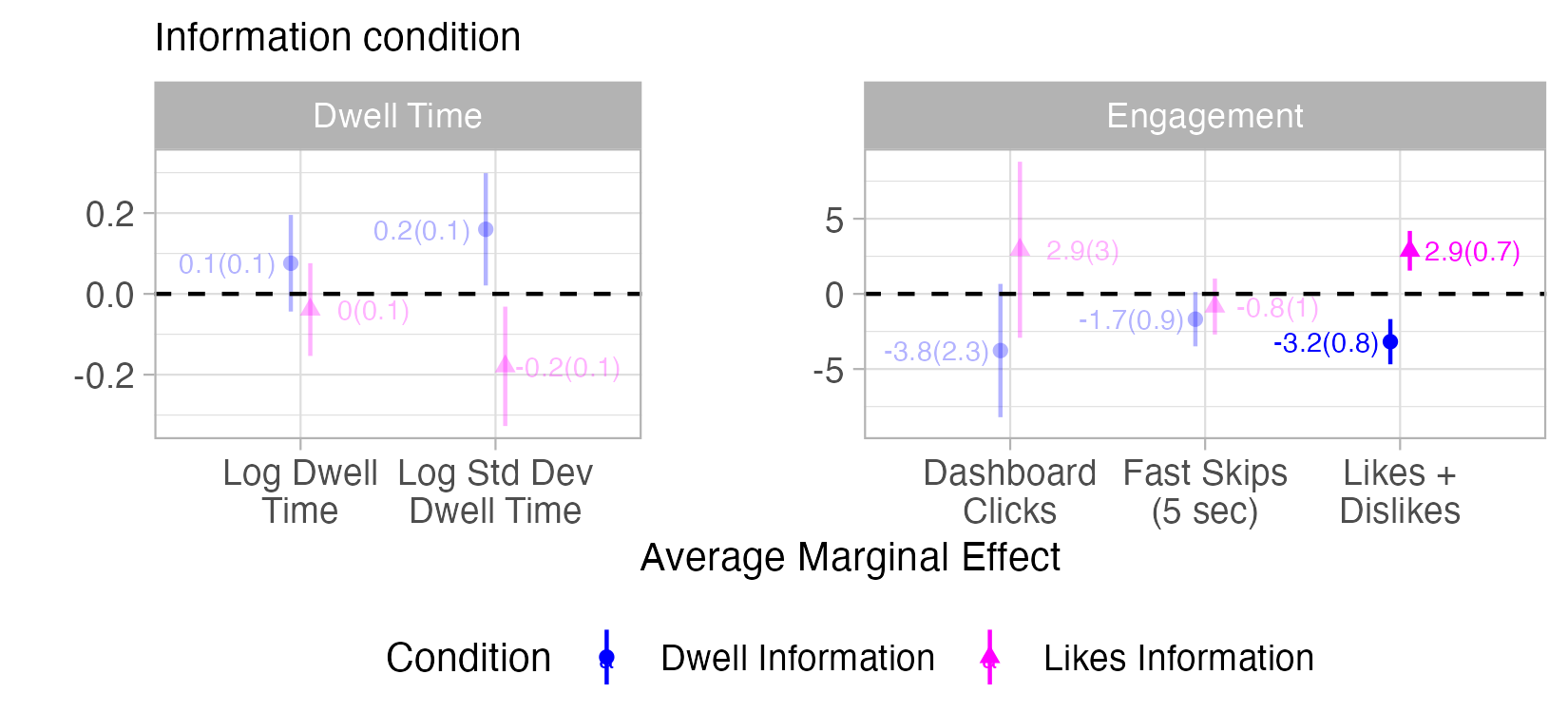}
        \caption{Effect of the ``Likes'' Information condition and the ``Dwell'' Information condition, compared to the ``Control'' Information condition, on participant behavior. \textit{Left:} Average marginal effects  of Information conditions on dwell time outcomes. Models are estimated using an OLS regression with controls for behavior in the Warm-up session. \textit{Right:}
        Average marginal effects of Information conditions on engagement outcomes. Models are estimated using a quasi-poisson regression with controls for behavior in the Warm-up session.}
\label{fig:information-main-effects}
\end{figure}

\subsubsection{Strategization across Information conditions.}
We first consider \textbf{H1}, which states that all else constant, providing participants with different descriptions of how their preferences are learned leads to different participant behaviors. We test this hypothesis by examining how participant behaviors differ across Information conditions, where participants are either told that the algorithm is  tracking (i) their ``likes'' and ``dislikes,'' which we refer to as the ``Likes'' Information condition; (ii) the time they spent listening to each song, which we refer to as the ``Dwell'' Information condition; or else are told (iii) no explicit information about what data the algorithm is tracking, which we refer to as the ``Control'' Information condition. 
Note that although we pre-registered models that include interactions, which are shown in Figure~\ref{fig:average-by-condition} and Appendix Table~\ref{appendix:additional-table-ate}, we do not find any significant interactions between the Information and Incentive conditions and thus consider the main effects of these conditions separately in the discussion below.

Overall, we find strong support for \textbf{H1}, 
as summarized in Figure~\ref{fig:information-main-effects}, 
which visualizes average marginal effects, 
pooled across the Incentive conditions. 
Participants' engagement patterns differ substantially across Information conditions and,as discussed in \cref{sec:discussion},
careful analysis suggests
experimenter demand effects are insufficient to explain the observed differences.
Below, we more closely inspect the effects of the Information condition on our various outcome metrics (pooling across different levels of the Incentive condition) and controlling for participants behavior in the Warm-up Session.
\\

\paragraph{Engagement metrics.}
As expected, the number of ``likes'' and ``dislikes'' (i.e., the Likes + Dislikes metric) is higher for participants who believe that the algorithm is tracking these actions more than others. Participants in the ``Likes'' Information condition generate 2.9 (SE: .7, $p <$ .001) more ``likes'' and ``dislikes'' on average than participants in the ``Control'' Information condition.

Interestingly, the number of “likes” and “dislikes” decreases when participants are in the “Dwell” Information condition. Participants in the “Dwell'' Information condition, who are told that the algorithm is paying attention to how long they spend listening to each song, submit 3.0 (SE: $.7$, $p < .001$) \emph{fewer} “likes” and “dislikes ”fewer dashboard clicks on average than those in the “Control'' Information condition. This finding suggests that our results are not merely reflecting experimenter demand – the “Dwell Time'' condition does not explicitly mention likes or dislikes – but due to the participants’ inferred model of the algorithm. Furthermore, that participants in the “Dwell'' condition submit fewer “likes” and “dislikes” “Control'' condition suggests that in the absence of explicit information in the “Control'' condition, participants make assumptions about how the algorithm is learning, e.g. due to their past experiences with music platforms like Spotify or Pandora. These results suggest that participants develop an understanding of how algorithms learn preferences and adjust their behavior based on this understanding, as we would expect based on our model of strategic behavior.

Overall, the effects of the Information treatment on ``likes'' and ``dislikes'' are substantial in magnitude. For example, participants in the ``Likes'' condition red (M=11.7) ``like'' or ``dislike'' ~80 percent more songs than participants in the ``Dwell'' condition (M=6.4). That is, with only minor differences in information about which data the platform is tracking, and \textit{no actual changes to the algorithm itself}, our treatment induces large differences in engagement behavior.
We observe the same (non-significant) trend in Dashboard Clicks but not in Fast Skips, as shown in Figure~\ref{fig:average-by-condition}. 
Since skips are related to dwell time (a greater number of skips implies a shorter dwell time for a fixed, five-minute listening session), 
we discuss skips next.

\paragraph{Dwell time metrics.}
While we do not observe significant average differences of the Information condition on the Fast Skips or the Log Average Dwell Time metrics of the participants,  
we do see weak evidence that the Information condition affects 
the variance of participants' dwell time. 
Specifically, participants in the ``Likes'' Information condition have a smaller Log Standard Deviation of Dwell Time than participants in the ``Control'' Information condition { ($\delta$ = -.2, $p = $ .06)}, and participants in the ``Dwell'' Information condition have a larger Log Standard Deviation of Dwell Time than participants in the ``Control'' Information condition { ($\delta$ = .2, $p =$ .06)}. 
The latter finding is consistent with our hypotheses that, if participants strategize with respect to dwell time, 
they would do so by emphasizing their relative interest in content (i.e., devote more dwell time on songs they want the algorithm to recommend and less on songs they want the algorithm to ignore), 
thus increasing the variance of dwell time across songs rather than the average dwell time itself. 
Notably, the ``Likes'' Information condition leads to a lower variance in dwell time compared to the ``Control'' Information condition, suggesting that participants are more strategic with respect to dwell time in the ``Control'' Information condition (in which they are told to behave as they would on Spotify or Pandora) than they are when they believe the algorithm primarily uses ``likes'' and ``dislikes'' to learn. 
While the effects for the Log Standard Deviation of Dwell Time are suggestive, we note that they are not significant at the $\alpha=.05$ level after adjusting for multiple comparisons.

\paragraph{Note on the ``Control'' Information condition.}
That participant behavior is different under the ``Control'' Information condition for outcome metrics that are \emph{not} cued in the instructions also suggests that participants come in with pre-conceived notions of how recommendation algorithms work. 
For example,  participants in the ``Dwell'' Information condition submit fewer ``likes'' and ``dislikes'' than those in the ``Control'' Information condition,
but (i) ``likes'' and ``dislikes'' are not mentioned in the ``Dwell'' or ``Control'' Information condition instructions
\emph{and} (ii) the ``Control'' Information condition asks participants to behave as they would on Spotify or Pandora. 
This difference indicates not only that participants already assume that recommendation algorithms on platforms like Spotify and Pandora pay attention to ``likes'' and ``dislikes,''
but also that their behavior reflects these assumptions. 
These results are consistent with an account that users strategize ``in the wild.''

\begin{figure}[t]
    \centering
    \includegraphics[width = 0.9\textwidth]{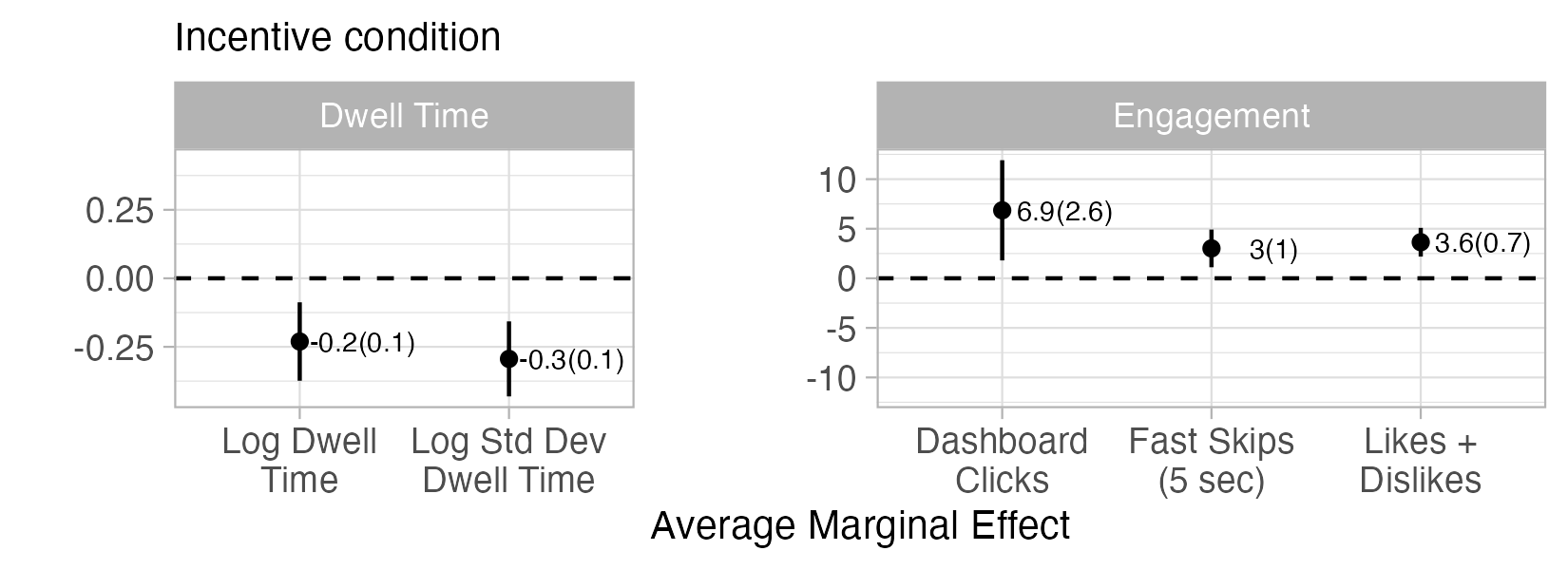}
        \caption{Effect of the Treatment Incentive condition compared to the Control Incentive condition on participant behavior. \textit{Left:} Average marginal effects of the Incentive condition on dwell time outcomes. Models are estimated using an OLS regression. \textit{Right:}
        Average marginal effects of Incentive  condition on engagement outcomes. Models are estimated using a quasi-poisson regression. Neither model is estimated with Warm-up session controls because participants are randomized into an Incentive condition before the Warm-up session.}
       
\label{fig:incentive-main-effects}
\end{figure}

\subsubsection{Strategization across Incentive conditions.}
We now turn to consider \textbf{H2}, which posits that participants who believe their behaviors affect future recommendations behave differently than participants who are not told they will receive recommendations at the end of the study.
We test this hypothesis by examining how participant behavior differs in different Incentive conditions, in which participants are either told that the algorithm is learning their preferences in order to provide them with personalized recommendations that they must evaluate after their listening sessions (the ``Treatment'' Incentive condition), or told that the algorithm is learning general music preferences (the ``Control'' Incentive condition). Analogously to the previous analysis, the reported results are estimated by pooling across different levels of the Information condition. Because participants are randomized to the Incentive condition before the Warm-up session, we do not include controls for participants' Warm-up session behavior (since doing so would be controlling for a post-treatment variable). 
Overall, we find strong support for \textbf{H2}: 
Participants' engagement patterns differ substantially across Incentive conditions, 
as summarized in Figure~\ref{fig:incentive-main-effects}. 
\\

\paragraph{Engagement metrics.}
Participants in the ``Treatment'' Incentive condition engage at higher rates than participants in the ``Control'' Incentive condition. The results are summarized in Figure~\ref{fig:incentive-main-effects}. Participants in the ``Treatment'' Incentive condition, who are told that the algorithm is tracking their behavior in order to provide them with personalized recommendations, submit 3.6 (SE: .7, $p <$ .001) more ``likes'' and ``dislikes,'' 3.0 more ``fast skips'' (SE: .97, $p$ = .009), and 6.9 more dashboard clicks (SE: 2.6, $p = $ .04) than participants in the ``Control'' Incentive condition. These results show that the participants in the ``Treatment'' Incentive  condition engage more with the music player than participants in the ``Control'' Incentive condition, 
presumably to shape their future recommendations. 
These effects, which are large in magnitude, as can be seen in Figure~\ref{fig:average-by-condition}, are evidence that participants take a more active role when they believe that their behavior affects their future recommendations. 

\paragraph{Dwell time metrics.}
Additionally, we find that participants in the ``Treatment'' Incentive condition listen to songs for a shorter amount of time  ($\delta$ = -.2, SE:.1, $p = $.006), and have a smaller standard deviation of dwell time ($\delta$ = -.3, SE: .1, $p <$ .001) than participants in the ``Control'' Incentive condition. 
{These results,} along with the higher number of  ``fast skips,'' shows that participants in the ``Treatment'' Incentive condition are skipping through songs relatively quickly on average, rather than exploring songs for longer amounts of time. 
{That participants in the ``Treatment'' Incentive have a lower standard deviation of dwell time aligns with the lower log dwell time 
since dwell time must be non-negative. 
Both trends are consistent with an account of participants who believe they are will receive personalized recommendations seeking to ``train'' the algorithm by sifting through content quickly to provide more feedback, %
rather than maximizing short-term utility by listening to songs they enjoy for longer periods of time. 
That participants who believe they will receive personalized recommendations, on average, increase explicit engagement (e.g., ``likes'' and ``dislikes'') and sift through songs faster rather than leverage dwell time variation is discussed next.}

\paragraph{Interaction between conditions.}
As can be seen in \cref{tab:olsTRUE,tab:proportionTRUE,tab:quasi-poissonTRUE}
and \cref{fig:average-by-condition}, we do not see strong evidence of an
interaction effect between our Information and Incentive conditions on any of
our outcome variables. This is surprising, since one might have expected that
participants in the ``Treatment'' Incentive condition might show intensified
effects of the various Information conditions---e.g., we might have expected a
positive interaction between the (i) ``Likes'' Information condition and
``Treatment'' Incentive condition on our Likes+Dislikes outcome or the (ii)
``Dwell'' Information condition and the ``Treatment'' Incentive condition on our
Log Standard Deviation Dwell Time outcome. Instead, we see that, on average, 
participants in the Incentive ``Treatment''
Condition (i) submit \emph{more} likes and dislikes and (ii) show \emph{smaller}
variation in dwell time than those in the ``Control'' Incentive condition. 
{The fact that
participants, when incentivized, primarily (i) increase engagement and (ii) do
not increase dwell time variation suggests that participants, on average, share
an implicit belief that recommendation algorithms pay more attention to
engagement than dwell time.} For example, participants might draw on past
experiences with music recommendation algorithms and assume that explicit
``likes'' is the most effective way to shape future recommendations {and, when
incentivized, increase this behavior on average}, regardless of the
Information condition. 

\paragraph{Experimenter demand effects.}
Experimenter demand is not sufficient to explain our results. 
Participants are given no indication that we track their dashboard clicks (across Information conditions, we only mention ``likes,'' ``dislikes,'' and ``dwell time''), 
but we observe a significant effect of the Incentive ``Treatment'' on dashboard clicks.
We even observe an effect of the ``Treatment'' Incentive condition in the ``Control'' Information  condition, in which participants receive \textit{no information about which behaviors we track} as is indicated by the red bars in Figure~\ref{fig:average-by-condition}, which show an increase in likes, fast skips, and dashboard clicks in the ``Treatment'' Incentive condition (compared to ``Control'' Incentive condition). 

\subsection{Subgroup Analysis}
We have thus far found evidence that users, on average, strategize when interacting with 
recommendation algorithms. 
In this section, we explore whether certain types of participants more likely to strategize than others. In particular, we highlight two theoretically relevant individual characteristics: Age and Spotify Use. (For brevity, we only examine a subset of our outcome variables and potential moderating characteristics; for the full set of moderators and outcomes, see Appendix~\ref{appendix:subgroup}). 
All the results below (and \cref{tab:dic_quasi-poisson_information_condition_with_controls_age,tab:dic_quasi-poisson_incentive_condition_without_controls_age,tab:dic_quasi-poisson_information_condition_with_controls_spotify_how_often,tab:dic_quasi-poisson_incentive_condition_without_controls_spotify_how_often}) give the CATEs and DICs by subgroup across Incentive and Information conditions.
Note that, when we refer to average treatment effects, 
we are computing the average marginal effect. 

\begin{table}

\caption{\label{tab:dic_quasi-poisson_incentive_condition_without_controls_age}Conditional Average Treatment Effects (CATEs) and Difference-in-CATEs (DIC) of the Incentive condition, by Age, Without Controls, pooled across Information Conditions}
\centering
\fontsize{9}{11}\selectfont
\begin{tabular}[t]{llll}
\toprule
\multicolumn{1}{c}{\bgroup\fontsize{10}{12}\selectfont \textbf{Outcome}\egroup{}} & \multicolumn{2}{c}{\bgroup\fontsize{10}{12}\selectfont \textbf{ATE}\egroup{}} & \multicolumn{1}{c}{\bgroup\fontsize{10}{12}\selectfont \textbf{DIC}\egroup{}} \\
\cmidrule{2-3}
\multicolumn{1}{c}{} & \multicolumn{1}{c}{Above-35} & \multicolumn{1}{c}{Below-35} & \multicolumn{1}{c}{}\\
\midrule
\multicolumn{1}{c}{Likes + Dislikes} & \multicolumn{1}{c}{4.4***(1.01)} & \multicolumn{1}{c}{2.65*(1.06)} & \multicolumn{1}{c}{-1.74(1.47)}\\
\multicolumn{1}{c}{Fast Skips (5 sec)} & \multicolumn{1}{c}{1.29(1.29)} & \multicolumn{1}{c}{5.64***(1.58)} & \multicolumn{1}{c}{4.36*(2.04)}\\
\multicolumn{1}{c}{Dashboard Clicks} & \multicolumn{1}{c}{12.64***(3.62)} & \multicolumn{1}{c}{2.72(4.32)} & \multicolumn{1}{c}{-9.91†(5.64)}\\
\bottomrule
\multicolumn{4}{l}{\textsuperscript{a} Heteroskedasticity Robust Standard Errors in Parentheses.}\\
\multicolumn{4}{l}{\textsuperscript{b} Signif. Codes: ***: .001, **: .01, *: .05, †: .1}\\
\end{tabular}
\end{table}

\begin{table}

\caption{\label{tab:dic_quasi-poisson_information_condition_with_controls_age}Conditional Average Treatment Effects (CATEs) and Difference-in-CATEs (DIC) of the Information condition, by Age, With Controls, pooled across Incentive conditions}
\centering
\fontsize{9}{11}\selectfont
\begin{tabular}[t]{lllll}
\toprule
\multicolumn{1}{c}{\bgroup\fontsize{10}{12}\selectfont \textbf{Outcome}\egroup{}} & \multicolumn{1}{c}{\bgroup\fontsize{10}{12}\selectfont \textbf{Info condition}\egroup{}} & \multicolumn{2}{c}{\bgroup\fontsize{10}{12}\selectfont \textbf{ATE}\egroup{}} & \multicolumn{1}{c}{\bgroup\fontsize{10}{12}\selectfont \textbf{DIC}\egroup{}} \\
\cmidrule{3-4}
\multicolumn{1}{c}{} & \multicolumn{1}{c}{} & \multicolumn{1}{c}{Above-35} & \multicolumn{1}{c}{Below-35} & \multicolumn{1}{c}{}\\
\midrule
\multicolumn{1}{c}{Likes + Dislikes} & \multicolumn{1}{c}{Likes} & \multicolumn{1}{c}{2.15**(0.78)} & \multicolumn{1}{c}{3.55**(1.09)} & \multicolumn{1}{c}{1.4(1.35)}\\
\multicolumn{1}{c}{Likes + Dislikes} & \multicolumn{1}{c}{Dwell} & \multicolumn{1}{c}{-3.66***(0.94)} & \multicolumn{1}{c}{-2.21*(1.09)} & \multicolumn{1}{c}{1.45(1.44)}\\
\multicolumn{1}{c}{Fast Skips (5 sec)} & \multicolumn{1}{c}{Likes} & \multicolumn{1}{c}{-0.48(1.17)} & \multicolumn{1}{c}{-2.1(1.5)} & \multicolumn{1}{c}{-1.62(1.91)}\\
\multicolumn{1}{c}{Fast Skips (5 sec)} & \multicolumn{1}{c}{Dwell} & \multicolumn{1}{c}{-1.5(1.11)} & \multicolumn{1}{c}{-1.81(1.34)} & \multicolumn{1}{c}{-0.31(1.74)}\\
\multicolumn{1}{c}{Dashboard Clicks} & \multicolumn{1}{c}{Likes} & \multicolumn{1}{c}{-1.31(3.58)} & \multicolumn{1}{c}{7.9†(4.56)} & \multicolumn{1}{c}{9.21(5.63)}\\
\multicolumn{1}{c}{Dashboard Clicks} & \multicolumn{1}{c}{Dwell} & \multicolumn{1}{c}{-3.78†(2.29)} & \multicolumn{1}{c}{-3.17(4.19)} & \multicolumn{1}{c}{0.61(4.78)}\\
\bottomrule
\multicolumn{5}{l}{\textsuperscript{a} Heteroskedasticity Robust Standard Errors in Parentheses.}\\
\multicolumn{5}{l}{\textsuperscript{b} Signif. Codes: ***: .001, **: .01, *: .05, †: .1}\\
\end{tabular}
\end{table}

\begin{figure}[t]
    \centering
    \includegraphics[width = 0.7\textwidth]{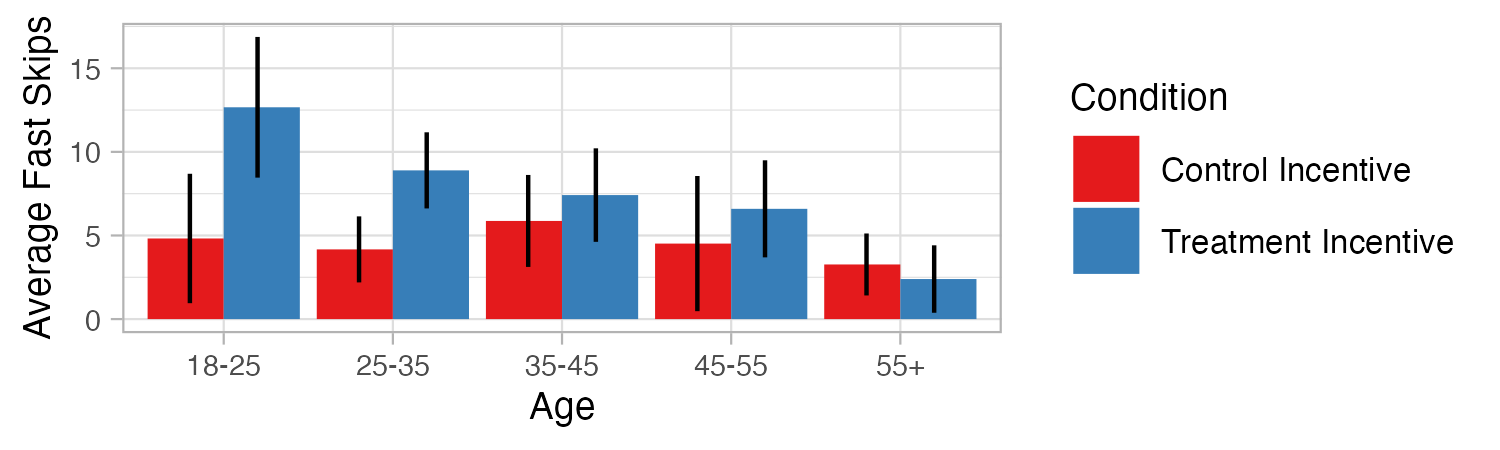}
    \caption{Average number of ``fast skips'' (with 95\% CIs) for ``Treatment'' vs. ``Control'' Incentive groups, for different Age groups }\label{fig:mean-incentive-effects-age}
\end{figure}

\subsubsection{Age.} 
We consider the participant's age as a potential moderator of our treatment effects. Research has shown that younger people are more familiar with technology and have higher digital literacy than older adults, with people born after 1980 sometimes referred to as ``digital natives'' \citep{palfrey_born_2011, vercruyssen2023basic, broady_comparison_2010}. Moreover, TikTok, a platform heavily dependent on recommendation algorithms for content curation and known from previous qualitative studies to have users who show awareness of the algorithm's behavior, is exceptionally popular among the younger demographic \citep{klug_trick_2021,anderson_social_2021}. 
Therefore, we hypothesize that younger participants would show more evidence of strategization. Specifically, we examine whether Age (split at 35 years old) moderates the effects of Incentive and Information conditions. 

Against our expectations, we do not find consistent evidence that younger participants are significantly more affected by our treatments (although we note that we are not sufficiently powered to detect small differences across age groups). The results are summarized in \cref{tab:dic_quasi-poisson_incentive_condition_without_controls_age}
and 
\cref{tab:dic_quasi-poisson_information_condition_with_controls_age}. 
As can be seen, participants below age 35 and above age 35 \textit{both} exhibit evidence of strategic behavior. For example, both older and younger participants show a significantly higher number of ``likes'' and ``dislikes'' in the ``Treatment'' Incentive condition and the ``Likes'' Information  condition, and show a significantly lower number of ``likes'' and ``dislikes'' in the ``Dwell'' Information condition. Furthermore, the difference in conditional average treatment effects (DICs) for older and younger participants is not significant across any measures (with one exception, which we consider in the next paragraph). These findings demonstrate that even older participants---who one might expect would show less sophistication when interacting with algorithms---exhibit evidence of strategic behavior on average.

A closer examination of the results reveals some evidence that the \textit{mechanism} of strategization---i.e., the differences in types of behaviors---differs for older vs. younger participants. In particular, we consider the Fast Skips metric---the number of times 
the participant skips past a song within 5 seconds of it starting. Figure~\ref{fig:mean-incentive-effects-age} shows a substantial trend between participants' age group and the average number of Fast Skips in the ``Treatment'' Incentive condition, where younger participants produce substantially more fast skips. As shown in Table~\ref{tab:dic_quasi-poisson_incentive_condition_without_controls_age}, participants below 35 have 5.64 more Fast Skips in the ``Treatment'' Incentive condition than ``Control'' Incentive condition, compared to only 1.29 more Fast Skips for participants above 35, for a DICs of 4.36 ($p=.03$). In contrast, participants above 35 show substantial evidence of strategization on our engagement metrics. Participants above 35 have 4.34 more likes and dislikes and 12.64 more dashboard clicks in the ``Treatment'' Incentive condition than in the ``Control'' Incentive condition, compared to 2.7 likes and dislikes and 2.7 dashboard clicks for participants under 35, respectively (although the DICs are not statistically significant).

\subsubsection{Spotify Use.} We next consider the participants' prior Spotify use as a moderator. We hypothesize that  participants with greater Spotify use would exhibit greater levels of strategization for two reasons. First, frequent participants are likely more familiar with music recommender platforms and, therefore, might be more comfortable in engaging with the platform. Second, frequent users of Spotify likely enjoy music and therefore might be more incentivized to strategize in order to receive higher payoff in the form of better song recommendations. 

As shown in Table~\ref{tab:dic_quasi-poisson_information_condition_with_controls_spotify_how_often}, we find evidence that both frequent and infrequent Spotify users respond to our Information conditions. However, as can be seen in Table~~\ref{tab:dic_quasi-poisson_incentive_condition_without_controls_spotify_how_often}, we do find some evidence that frequent Spotify users engage at higher rates in the Incentive condition. Frequent Spotify users (i.e. those who use the app more than once a week) submitted 4.68 more ``likes'' and ``dislikes,'' 4.17 more ``fast skips,'' and 9.83 more ``dashboard clicks'' in the Treatment vs. Control Incentive conditions. In contrast, less frequent users submitted 1.54 more ``likes'' and ``dislikes,'' .87 more Fast Skips, and .13 more ``dashboard clicks'' in the ``Treatment'' vs. ``Control'' Incentive condition. These give us DICs of 3.13 ($p=.03$), 3.30 ($p=.09$), 9.7 ($p=.08$), respectively, showing that the effect of the ``Treatment'' Incentive condition is larger for participants who use Spotify more frequently. We note that while these differences are large in magnitude, they are imprecisely measured. This is because our analysis is underpowered to detect small differences across groups, largely because the relatively few number of infrequent Spotify users in our experiment (200 infrequent vs. 452 frequent users). Nonetheless, these findings are suggestive: 
participants who we would expect to derive more long-term payoff from better song recommendations exhibit greater evidence of strategization in the Incentive ``Treatment.''
\\

\noindent
In summary, while we find evidence that the degree and mechanism of strategization might vary across subgroups, we do not find evidence that strategic behavior is wholly concentrated within a particular demographic. Rather, some degree of strategization is common across types of users. 
Additionally, we see some evidence that the level of strategization is  related to the perception of the long-term payoff. Participants who we would expect to place more value on the future song recommendations (based on how often they use music streaming platforms) show more evidence of strategization. 
Deeper analysis examining how strategic behavior varies across different types of users is an important area of future work for researchers and platforms. 

\begin{table}

\caption{\label{tab:dic_quasi-poisson_incentive_condition_without_controls_spotify_how_often}Conditional Average Treatment Effects (CATEs) and Difference-in-CATEs (DIC) of the Incentive condition, by Spotify Use, Without Controls, pooled across Information Conditions}
\centering
\fontsize{9}{11}\selectfont
\begin{tabular}[t]{llll}
\toprule
\multicolumn{1}{c}{\bgroup\fontsize{10}{12}\selectfont \textbf{Outcome}\egroup{}} & \multicolumn{2}{c}{\bgroup\fontsize{10}{12}\selectfont \textbf{ATE}\egroup{}} & \multicolumn{1}{c}{\bgroup\fontsize{10}{12}\selectfont \textbf{DIC}\egroup{}} \\
\cmidrule{2-3}
\multicolumn{1}{c}{} & \multicolumn{1}{c}{Spotify Use=Often} & \multicolumn{1}{c}{Spotify Use=Rare} & \multicolumn{1}{c}{}\\
\midrule
\multicolumn{1}{c}{Likes + Dislikes} & \multicolumn{1}{c}{4.67***(0.94)} & \multicolumn{1}{c}{1.54(1.13)} & \multicolumn{1}{c}{3.13*(1.47)}\\
\multicolumn{1}{c}{Fast Skips (5 sec)} & \multicolumn{1}{c}{4.18**(1.27)} & \multicolumn{1}{c}{0.87(1.46)} & \multicolumn{1}{c}{3.31†(1.93)}\\
\multicolumn{1}{c}{Dashboard Clicks} & \multicolumn{1}{c}{9.83**(3.32)} & \multicolumn{1}{c}{0.13(4.35)} & \multicolumn{1}{c}{9.7†(5.47)}\\
\bottomrule
\multicolumn{4}{l}{\textsuperscript{a} Heteroskedasticity Robust Standard Errors in Parentheses.}\\
\multicolumn{4}{l}{\textsuperscript{b} Signif. Codes: ***: .001, **: .01, *: .05, †: .1}\\
\end{tabular}
\end{table}

\begin{table}

\caption{\label{tab:dic_quasi-poisson_information_condition_with_controls_spotify_how_often}Conditional Average Treatment Effects (CATEs) and Difference-in-CATEs (DIC) of the Information condition, by Spotify Use, With Controls, pooled across Incentive conditions}
\centering
\fontsize{9}{11}\selectfont
\begin{tabular}[t]{lllll}
\toprule
\multicolumn{1}{c}{\bgroup\fontsize{10}{12}\selectfont \textbf{Outcome}\egroup{}} & \multicolumn{1}{c}{\bgroup\fontsize{10}{12}\selectfont \textbf{Info condition}\egroup{}} & \multicolumn{2}{c}{\bgroup\fontsize{10}{12}\selectfont \textbf{ATE}\egroup{}} & \multicolumn{1}{c}{\bgroup\fontsize{10}{12}\selectfont \textbf{DIC}\egroup{}} \\
\cmidrule{3-4}
\multicolumn{1}{c}{} & \multicolumn{1}{c}{} & \multicolumn{1}{c}{Spotify Use=Often} & \multicolumn{1}{c}{Spotify Use=Rare} & \multicolumn{1}{c}{}\\
\midrule
\multicolumn{1}{c}{Likes + Dislikes} & \multicolumn{1}{c}{Likes} & \multicolumn{1}{c}{3.05***(0.82)} & \multicolumn{1}{c}{2.28*(1.09)} & \multicolumn{1}{c}{0.77(1.39)}\\
\multicolumn{1}{c}{Likes + Dislikes} & \multicolumn{1}{c}{Dwell} & \multicolumn{1}{c}{-2.28*(0.89)} & \multicolumn{1}{c}{-4.88***(1.31)} & \multicolumn{1}{c}{2.6†(1.57)}\\
\multicolumn{1}{c}{Fast Skips (5 sec)} & \multicolumn{1}{c}{Likes} & \multicolumn{1}{c}{-0.36(1.23)} & \multicolumn{1}{c}{-2.47†(1.49)} & \multicolumn{1}{c}{2.1(1.94)}\\
\multicolumn{1}{c}{Fast Skips (5 sec)} & \multicolumn{1}{c}{Dwell} & \multicolumn{1}{c}{-1.43(1.17)} & \multicolumn{1}{c}{-1.47(1.3)} & \multicolumn{1}{c}{0.04(1.75)}\\
\multicolumn{1}{c}{Dashboard Clicks} & \multicolumn{1}{c}{Likes} & \multicolumn{1}{c}{1.73(4.39)} & \multicolumn{1}{c}{4.05(3.7)} & \multicolumn{1}{c}{-2.32(5.77)}\\
\multicolumn{1}{c}{Dashboard Clicks} & \multicolumn{1}{c}{Dwell} & \multicolumn{1}{c}{-3.33(2.89)} & \multicolumn{1}{c}{-6.13(3.78)} & \multicolumn{1}{c}{2.79(4.76)}\\
\bottomrule
\multicolumn{5}{l}{\textsuperscript{a} Heteroskedasticity Robust Standard Errors in Parentheses.}\\
\multicolumn{5}{l}{\textsuperscript{b} Signif. Codes: ***: .001, **: .01, *: .05, †: .1}\\
\end{tabular}
\end{table}

\subsection{Post-Experiment Survey}\label{sec:survey}

Our analysis of the experiment results in previous sections suggests that users strategize in response to Information and Incentives, as hypothesized. 
However, 
because our analysis surfaces \emph{average} effects, 
it does not determine \emph{how many} participants strategize. 
It is also unclear
whether participants are conscious of their strategization (either during the experiment or ``in the wild'') and, if they are, their reasons for strategizing.
In this section, we analyze two questions in the post-experiment survey. 
The first asks our participants whether they changed their behavior across listening sessions (i.e., strategized).
The second asks participants whether they strategize on their own platforms.
We find that many users self-report strategizing during the experiment and that users in the ``Likes'' Information condition and ``Dwell''  Information condition self-report higher levels of strategization. We further find that users self-report strategizing on their own platforms and present five qualitative themes in the responses for their motivation to do so.

\subsubsection{Self-reported changes in study behavior.}

\begin{figure*}[t]
        \centering
        \includegraphics[width=\textwidth]{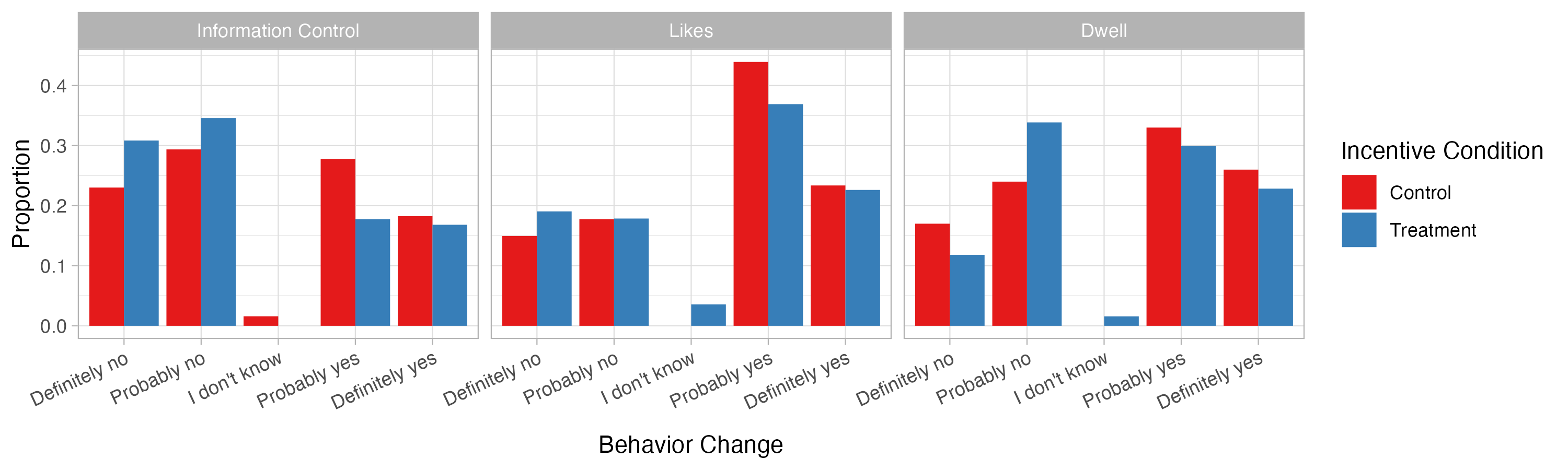}
        \caption{Proportion of participant responses to the post-experiment question ``Did the way that you interacted with songs change across the [...] listening sessions?'' 
        The results are grouped by Information and Incentive conditions, 
        which are separately analyzed in \cref{tab:change_interaction_incentive_info}. \hlcomment{Can you increase the space between the answer options?}}
        \label{fig:postexperiment_survey_Q1_full_results}
\end{figure*}

The first post-experiment survey question asks users to self-report whether they behave differently across listening sessions. Figure \ref{fig:postexperiment_survey_Q1_full_results} and Table~\ref{tab:change_interaction_incentive_info}
summarize the results. 
53 percent of participants state that they ``Probably'' or ``Definitely'' changed their behavior across sessions, suggesting a high level of conscious strategization in our experiment. Participants in the ``Likes'' and ``Dwell'' Information conditions are, respectively, 23 and 16 percentage points more likely to answer that they ``Probably'' or ``Definitely'' changed their behavior. This is consistent with an account in which participants in these conditions consciously adapt their behavior when receiving new information about what behaviors the algorithm tracks after the Warm-up session at the start of the listening session.

Notably, 
responses to this question do not definitively confirm or deny strategization, 
as the question leaves room for both under-reporting and over-reporting of strategization. 
On one hand, we may {\em under-estimate} strategization, since
self-reported changes in study behavior may not include all forms of (even conscious) strategization. 
For example, a user who is strategic across both the Warm-up and Listening sessions may report that they did not change their behavior.
Potentially illustrating this phenomenon, participants in the ``Treatment'' Incentive condition are 7 percentage points less likely to state behavior changes, though this effect is not statistically significant and is somewhat attenuated in the ``Likes'' and ``Dwell'' Information groups, 
as shown by the positive coefficients on the interaction terms in Table~\ref{tab:change_interaction_incentive_info}.
This lack of effect is not particularly surprising given that participants are randomized into the Incentive condition before the Warm-up session, unlike in the Information condition, and thus participants might adopt the same (potentially strategic) behavior beginning in the Warm-up session and kept it across all sessions.

On the other hand, the wording of our question may be picking up on behavior changes due to factors \emph{other than} strategization  and therefore over-estimate strategization, as evidenced by  the red bars on the left-most plot in \cref{fig:postexperiment_survey_Q1_full_results}, 
which show that 
some participants who receive both the Information ``Control'' and Incentive ``Control'' report changing their behavior across sessions. 
This could be due to the ambiguity of our question, where participants might change their behavior as they become acquainted with our platform, or due to some baseline level of strategization as participants moved from the ``Warm-up'' to the ``Listening'' session that was not induced by our experimental conditions.

We note that there are two key differences between our experiment and survey findings. 
First, the survey responses highlight the degree to which participants \emph{consciously} strategize, 
Second, since the analysis of the experiment results relies on average effects, 
we are unable to determine how many participants strategize. 
The post-experiment survey studied next provides insights into the rates of (conscious) strategization.

\begin{table}[htbp]
   \caption{\label{tab:change_interaction_incentive_info} OLS Regression for linear probability model predicting stated interaction change from experimental conditions}
   \centering
   \small
   \begin{tabular}{lc}
      \tabularnewline \midrule \midrule
      Dependent Variable:                  & Changed Interaction\\  
      Model:                               & (1)\\  
      \midrule
      \emph{Variables}\\
      Constant                             & 0.60$^{***}$ (0.02)\\   
      1(Incentive)                         & -0.07 (0.05)\\   
      1(Likes Info)                        & 0.23$^{***}$ (0.05)\\   
      1(Dwell Info)                        & 0.16$^{***}$ (0.05)\\   
      1(Incentive) $\times$ 1(Likes Info)  & 0.04 (0.10)\\   
      1(Incentive) $\times$ 1(Dwell Info)  & 0.05 (0.09)\\   
      \midrule
      \emph{Fit statistics}\\
      Observations                         & 651\\  
      R$^2$                                & 0.04\\  
      Adjusted R$^2$                       & 0.04\\  
      \midrule \midrule
      \multicolumn{2}{l}{\emph{Heteroskedasticity-robust standard-errors in parentheses}}\\
      \multicolumn{2}{l}{\emph{Signif. Codes: ***: 0.01, **: 0.05, *: 0.1}}\\
   \end{tabular}
   \vspace{10pt}
   \par \raggedright 
   {\em Experimental conditions are effects coded (i.e., centered dummy variables). The dependent variable ``Changed Interaction'' is coded as 1 if the participant answered ``Definitely Yes'' or ``Probably Yes'' to the question ``Did the way that you interacted with songs change across the three listening sessions?'', and ``0'' if the participant answered ``Don't know'', ``Probably No'', or ``Definitely No.''}
\end{table}

\subsubsection{Self-reported strategization ``in the wild.''}
In this section, 
we discuss participant responses to a free-form question about whether they strategize on their own platforms (i.e., outside of our study).
Specifically, we asked participants whether they try to ``talk'' to their algorithm or ``hide'' things from it
(the specific wording of post-experiment questions are given in \cref{app:survey_Qs}).
We provide an exploratory analysis, 
and we note that the results are not necessarily representative of the general population. 
For more information on the demographic split of our study, 
see \cref{app:demographic_plots}.
Although our analysis may not be representative, 
we believe it suggests that many users strategize ``in the wild'' and provides some insights into why and how they do so.

\heading{Summary statistics.}
We code the responses into three categories: 
definitive strategization, no strategization, 
and unclear. 
A response is classified as showing definitive strategization if the participant clearly demonstrates (1) that they anticipate how their current actions affect future recommendations
and (2) that this anticipation leads the participant to change their interaction behavior. 
A response is classified as not showing strategization if the participant does not indicate any strategization as defined above. 
To avoid over-reporting strategization, 
any responses that do not exhibit (1) and (2) simultaneously are categorized as unclear. 

We obtained summary statistics 
using ChatGPT, 
which we asked to classify the responses as described above in 10 separate sessions. 
The results are summarized in \cref{tab:talk_survey_summary_stats}.
These results show that around 47 percent of participants self-report to strategize ``in the wild.''

\heading{Qualitative analysis.}
Finally, we went through the responses by hand and analyzed why and how users strategize. 
We identified several trends that persist across participants and include example responses below.

\paragraph{Being pigeonholed by algorithms:}
Some participants express that they do not like to be pigeonholed by their algorithm,
with one stating
``what I like today might not be what I will like tomorrow,''
another saying ``Yes sometimes I may like a song but not thumbs-up the song because I don't want my feed filled with similar artists/videos.  This is because I might like only one type of song by an artist,''
and a third sharing that
``On YouTube I will like things I don't and dislike things I do and subscribe to dozens and dozens of channels, even walk out of the room with something I like or dislike playing just so I get lots of new stuff and they don't pigeonhole me too much and show me crap I don't want to see over and over. Basically, I try to be purposefully unpredictable and then go into my subscriptions and play from there the stuff I really want to see. My hope is the two are playing against each other and the algorithm doesn't know exactly what I want.''
Similarly,
several say that they like to reset their algorithm, 
stating: 
``I have played some music I would not normally listen to or even like to throw off an algorithm''
and 
``I might give thumbs-up to specific songs if I am trying to reset the algorithm and get it to forget what I have been listening to.''

\paragraph{Helping the algorithm.}
Several participants suggest that they strategize to help their algorithm identify their preferences. 
One said:
``[D]uring the sessions a Blink-182 song came on, and I'm not really crazy about them, but I was hoping to force the algorithm to swing more towards a `rock' vibe'' while another responded:
``Yes. Thumbs upped songs in this survey that I didn't like because I wanted to hear similar bands. I hated that Blink 182 song, but I love Blink and I love punk music so I thumbs upped it anyway. Sometimes you gotta play along with the algorithm if you want it to work best for you.''
Others said
``If I'm looking for more recommendations that are similar to a certain genre I will leave a playlist based on that genre playing for a day or two to try and get different recommendations matching those songs'' 
and 
``I have frequently given thumbs up or not skipped a mediocre song by an artist that I otherwise love because I want their songs to continue to show up.''
Some even indicate awareness of more subtle recommendation tactics, like dwell-time tracking: 
``If I see something that I know I am not interested in, I quickly click away from it, I do not want to linger too long or the algorithm may think I am interested and show me more like it.''

\begin{table}[t]
\centering
\begin{tabular}{lccc}
\hline
\textbf{Strategization} %
& \textbf{Average} & \textbf{Std Dev} \\ \hline
Yes             %
& 0.471   &      0.190              \\
No              %
&  0.283  &       0.159              \\
Unclear         %
&    0.247  & 0.212                   \\ \hline
\end{tabular}
\caption{ChatGPT analysis summary statistics of self-reported strategization in post-experiment survey}
\label{tab:talk_survey_summary_stats}
\end{table}

\paragraph{Preserving accounts.}
We find that several participants did not want to ``ruin'' their algorithm with unintended interactions:
``I try not to link my account to others to avoid them ``poisoning'' my algorithm with their preferences since algorithms assume there must be some kind of overlap between you and those you associate with'' and 
``Yes, I often do like songs or avoid clicking links or ads that would impact my user profile on various platforms. I am aware that my activity often gets tracked and that the algorithms on social media or music sites detect the changes and cater to my new preferences. Sometimes, I do not want that to happen so I avoid clicking links. If I am with a friend who has a different music taste and wants to search something on my phone, I am often scared that it will impact my own music recommendations and so I try to limit that.''
Several even confess to creating multiple accounts: ``“I have many YouTube accounts so my algorithm does not pick up  a YouTube link a friend sends me to watch.''

\paragraph{Private browsing.}
Many of our participants admit to using Incognito or private browsing mode to interact with interesting content:
``If I want to just check something but not mess up my preferences, I will use incognito mode in Chrome so I'm not signed in''
and 
``I avoid searching something embarrassing unless it is in incognito mode, because I expect I would get ads related to it after.''

\paragraph{Using tracking to their advantage.}
Some participants strategize off-platform, as epitomized by the response:
``If there's something I am interested in and haven't seen an ad for it, I will google it because I know within a very short amount of time, ads will start appearing in my feeds.''

\paragraph{No strategization.}
Many participants reported not strategizing, 
responding: 
``I believe that algorithms are a useful tool that can help us make better decisions and find new insights''
and
``I'm pretty (and blatantly) honest about my feelings. And yes, this sometimes gets me into trouble, but it's easier to be honest about something than not.''

\section{Discussion}\label{sec:discussion}

In this section, 
we review our main findings and their implications before discussing the main limitations of our study and directions for future work. 

\subsection{Summary of Findings}
Our results demonstrate that users are not only cognizant of recommendation algorithms, 
but also adapt their behavior based on knowledge of their algorithm. 
In other words, users {\em strategize} \citep{cen2023trust}.
While much of the discussion around recommendation focuses on the platform's role in shaping what users see or, in the case of social media, the content creator's role in choosing the type of content to create, we show that users also play an active role in shaping what they see.

We provide a first step into documenting and measuring strategization through an online lab experiment and survey. 
From our lab experiment, 
we find strong evidence of strategization: that participants adapt their behaviors to the (perceived) algorithm when they believe doing so elicits better recommendations (\cref{sec:results}).
In particular, %
we find that participants exposed to different information about (i) how the algorithm works and (ii) whether they will receive personalized recommendations use significantly different behaviors. 
The magnitude of this strategization is substantial and apparent across multiple outcome metrics.
Interestingly, we see significant behavioral differences even when we make very minor wording changes. 
Moreover, strategization is not just concentrated among particularly active users; for example, even users thought to be naive (e.g., older participants) exhibit significant evidence of strategization.

What is particularly notable is that we see evidence of strategization  
(i) even if we consider experimenter demand effects and 
(ii) despite possible spillover from participants' pre-existing notions about how popular recommendation algorithms work. 
For a discussion of demand effects in the context of our experiment, 
see ``Limitations'' below. 
With respect to spillover, 
recall that our Information ``Control'' asks that participants behave as they would on their music platform of choice.
Since users may already strategize on their music platforms, 
our ability to detect significant effects under the Information ``Treatments'' (both of which overlap with how popular algorithms actually work) strengthens our findings rather than weakens them.

In our post-experiment survey, 
we uncover \emph{why} and \emph{how} participants strategize ``in the wild'' (i.e., when interaction with their own platforms). 
Our participant responses reveal that users strategize both to help the algorithm and to obscure information from the algorithm, among other reasons. 
We also find that around 47 percent of users report definitive strategization ``in the wild'' while 28 percent report that they do not strategize, 
and 25 percent provide answers that are ambiguous.

\subsection{Managerial Implications}
Our study is part of a larger body of work that suggests that engagement is not always an accurate proxy for utility, and that platforms must be careful when training recommendation algorithms on engagement. 
For example, other works explore how users behave differently when making ``snap'' versus ``deliberate'' decision \citep{agan2023automating} 
and how users may behave differently under ``inconsistent preferences'' \citep{Kleinberg2022-wy}.
Within this growing literature, 
our work explores whether user behavior is \emph{algorithm-dependent} (whereas the two phenomena above can occur under the same algorithm). 
Although we study recommendation, 
our findings may extend to other contexts in which users are aware of their algorithm (e.g., Uber or TaskRabbit). 

That user behavior can be algorithm-dependent has substantial managerial implications about the interpretability of the data \citep{cen2023trust}. 
It implies, for example, that behaviors under different algorithms may not be directly comparable.
Consider a case where a platform tests a new recommendation algorithm that increases the amount of user engagement (e.g., clicks) on recommended content. Common wisdom would suggest that the platform should roll out the new algorithm, since users seem to prefer the content from the algorithm, due to the increase in clicks. However, it is possible that users actually like the suggestions by the new algorithm \textit{less} and are clicking ``like'' more often in order to nudge the algorithm to make \textit{different} recommendations. 
In this way, 
algorithm-dependent user behaviors are not always comparable and may even be misinterpreted. %

Interestingly, our survey results (\cref{sec:survey}) suggest that users strategize both to ``help'' and ``hurt'' the platform. 
In some cases, users strategize to better \textit{communicate} to their preferences to the platform based on their understanding of their algorithm. 
In others, they adapt their behaviors to ``obscure'' information from the algorithm. 
This ambiguity further indicates that behavioral data suffers from an interpretability problem. 

We also emphasize that our experiment shows that even \textit{small} wording changes (and no algorithmic changes) can induce \textit{significant} effects. Thus it is possible that even small algorithmic or user interace changes in the recommendation system can influence users' perception about the algorithm and change the nature of the data gathered, 
as hypothesized by \citet{cen2023trust}.

\subsection{Limitations}
Next, we consider potential limitations of our work: experimenter demand effects, 
external validity, 
and our focus on average effects. 

Demand effects arise when cues given by the experimenter (unintentionally) reveals how the experimenter expects participants to behave, 
and these cues influence participants' behaviors. 
In our experiment, 
demand effects could surface in the ``Likes'' and ``Dwell'' Information conditions,
in that users may (i) change their ``likes'' and ``dislikes'' behavior or (ii) move through songs more quickly in response to the ``Likes'' and ``Dwell'' Information conditions, respectively.
However, we highlight three observations that 
{suggest that our results are robust to demand effects}: 
\begin{enumerate}
    \item Although the Information conditions might induce a demand effect, it is very unlikely the Incentive conditions do because each participant is only assigned to one Incentive condition (making it difficult for participants to infer the other  condition), 
and the Incentive condition does not reveal information about the algorithm's inner workings. 
Yet we see statistically significant effects from the 
``Treatment'' Incentive condition across all outcome metrics (\cref{fig:incentive-main-effects}). 
    \item Even for the Information condition, 
we observe statistically significant differences between Information conditions for behaviors that are \emph{not} mentioned to participants and therefore are unlikely to be the result of demand effects.  
For example, participants in the ``Dwell'' Information group significantly change their ``likes'' behavior (\cref{fig:information-main-effects}),
and
participants in the ``Likes'' Information group significantly change their ``dwell time'' variation.
    \item We see significant heterogeneity in the method of strategization by age (Fig. \ref{fig:mean-incentive-effects-age})%
    . For example, younger participants used more ``fast skips'' and fewer ``likes'' in the ``Treatment'' Incentive condition, and older participants did the opposite. Such heterogeneity is unlikely to be driven by a demand effect since our experimental conditions are independent of age.
\end{enumerate}

One of the main limitations of our work involves the question of external validity: how accurately our findings reflect real-world behaviors. 
Our post-experiment survey helps to close this gap by asking whether users strategize ``in the wild,'' to which 
almost half of participants definitively report strategizing (\cref{sec:survey}). 
Even so, 
the extent to which our results are externally valid remains unclear in two ways. 
First, 
in order to decouple our results from participants' pre-existing beliefs about recommendation algorithms (which can be highly heterogeneous and therefore reduce experimental power), 
we explicitly reveal information about the algorithm to users (via the Information conditions). 
In most real-world platforms, 
users are not given explicit information about their algorithm's inner workings; instead, users learn how their algorithm works through repeated interaction. 
Although we consider the question of how accurately users are able to infer their algorithms' inner workings to be out-of-scope (see, for example, \citet{martens2023decoding}), 
this is an inherent limitation of our work. 
Second, 
we experiment with music recommendation because it is a fairly universal setting and therefore appropriate for a first empirical test of strategization.%
However, it is unclear whether and how our findings would extend to other contexts,
especially higher-stakes ones, such as news recommendation or labor marketplaces. 
We hypothesize that strategization is even more prevalent on platforms like TikTok, 
where awareness of the algorithm is stronger than in the music recommendation setting. 
Therefore, although our experiment design allows us to control the environment and draw clean conclusions,
external validity is an unavoidable consideration. 

Another limitation of our work is our focus on average treatment effects. 
Under our experiment design, 
drawing conclusions about individual treatment effects is difficult. 
For example, our analysis does not shed light on how many participants strategized (since we cannot determine how participants \emph{would have} behaved in different conditions without spillover effects). 
For the same reason, we cannot determine the behaviors individual users leverage when strategizing.\footnote{Interestingly, heterogeneity in how users strategize strengthens our results because we detect statistically significant average treatment effects even though some participants may behave ``oppositely,'' i.e., one participant's behavior change may ``cancel out'' another's).}

\subsection{Future Work}
The limitations described above suggest directions for future work, which we explore next.

To test the external validity of our findings,
a natural direction for future work is running field experiments or studies that measure and characterize user strategization on real-world platforms (including those that are not traditionally considered recommendation platforms but are powered by algorithms trained on user behavior). 
Comparing the level and nature of strategization across data-driven platforms would help to understand, in broad strokes, (i) the factors that elicit strategization and (ii) the mechanisms that users leverage when strategizing. 
While \citet{cen2023trust} catalog various reasons why users might strategize, there is a lack of empirical evidence on the relative importance of these reasons for users. 

Our post-experiment study suggests both why and how users strategize in the wild. 
For example, some participants report that they use different accounts to access different types of content (perhaps due to inconsistent preferences \cite{Kleinberg2022-wy}); 
engage with items that they do not like in order to encourage related content that they do like; 
avoid clicking on certain advertisements and posts to prevent repetitive content; 
and selectively browse in incognito mode to ``hide'' information from the algorithm. 
As such, one particular direction for future work would be to study the frequency of motivations and of methods for strategization, 
especially across different demographic groups. 
While some insights can be surfaced through an analysis of average treatment effect, 
the ideal would be to estimate individual treatment effects (i.e., estimate how a given user would have behaved under a different algorithm).
It would also be interesting to explore whether strategizing actually improves users' utilities.

Another way to test the external validity of our results is by allowing users to organically develop ``beliefs'' about how their algorithms work through repeated interaction.
Specifically, in our lab experiment, we measure strategization by influencing users' beliefs about the algorithm via the Information conditions; but, in practice, users simultaneously develop beliefs about their algorithm while strategically adapt their behaviors.
Thus understanding strategization ``in the wild'' must consider the heterogeneity of user experiences and, as a result, beliefs about their algorithm.

Finally, in the face of strategization, 
platforms must find more robust preference elicitation methods. 
Using an understanding of why and how users strategize, 
future work could test the effectiveness different methods (e.g., \citet{cen2023trust} suggest that participatory methods can improve  ``cooperation'' between users and the algorithm).

\newpage
\bibliographystyle{informs2014} %

\bibliography{ref,Empirical-Trust}

\newpage
\begin{APPENDICES}

\section{Additional Experimental Details}\label{app:additional_experimental_details}

\subsection{Covariate Balance}\label{appendix:balance-check}

In Table~\ref{tab:balance-check}, we report covariates across condition. We also report Chi-Squared tests to assess covariate balance and find that our demographics are balanced across conditions. 

\begin{table}

\caption{\label{tab:balance-check}Percent of Condition that (i) Is Female, (ii) Is Under 35, (iii) Is College Educated, (iv) Is Privacy Concerned, (v) Frequently uses social media, and (vi) Frequently uses Spotify Balance Checks for Demographics, by Conditions. Balance check is done via chi-square test, with p-values adjusted with a Benjamini-Hochbert correction.}
\centering
\resizebox{\linewidth}{!}{
\begin{tabular}[h]{l>{\raggedleft\arraybackslash}p{2.1cm}>{\raggedleft\arraybackslash}p{2.1cm}>{\raggedleft\arraybackslash}p{2.1cm}>{\raggedleft\arraybackslash}p{2.1cm}>{\raggedleft\arraybackslash}p{2.1cm}>{\raggedleft\arraybackslash}p{2.1cm}rr}
\toprule
Variable & Control Incentive
Control Info & Control Incentive
Dwell Info & Control Incentive
Likes Info & Treatment Incentive
Control Info & Treatment Incentive
Dwell Info & Treatment Incentive
Likes Info & statistic & p.adj\\
\midrule
Is Female & 50\% & 43\% & 53\% & 48\% & 55\% & 59\% & 6.44 & 0.53\\
Is Under 35 & 50\% & 43\% & 52\% & 36\% & 41\% & 40\% & 8.80 & 0.47\\
Has College Edu & 59\% & 57\% & 56\% & 69\% & 62\% & 51\% & 7.99 & 0.47\\
Is Privacy Concerned & 29\% & 32\% & 24\% & 30\% & 31\% & 31\% & 1.85 & 0.87\\
Is Frequent Social Media User & 75\% & 75\% & 77\% & 73\% & 68\% & 74\% & 3.13 & 0.87\\
\addlinespace
Is Frequent Spotify User & 42\% & 36\% & 40\% & 41\% & 36\% & 35\% & 2.30 & 0.87\\
\bottomrule
\end{tabular}}
\end{table}

\subsection{Models}
Below, we specify the models estimated for our main analyses. For each individual $i$, we have the following:
\begin{itemize}
    \item $\textbf{Y}_i$ is the value of $i$'s dependent variable of interest in the treatment session (number of likes + dislikes, number of skips, number of dashboard clicks, average song dwell time, and standard deviation of dwell time, respectively)
    \item $\textbf{Y}^0_i$ is the value of $i$'s variable of interest in the \textit{warmup} session (number of likes + dislikes, number of skips, number of dashboard clicks, average song dwell time, and standard deviation of dwell time, respectively)
    \item $\textbf{incent}_i$ is a centered dummy variable equal to .5 if the participant is assigned to the incentive treatment condition, else -.5
    \item $\textbf{likes}_i$ is a centered dummy variable equal to .5 if the participant is assigned to the likes information condition, else -.5
    \item $\textbf{dwell}_i$ is a centered dummy variable equal to .5 if the participant is assigned to the duration information condition, else -.5
\end{itemize} 

\paragraph{Count Models}\label{sec:model-count}
For our three count variables, i) Number of likes and dislikes, ii) Number of fast skips, and iii) Number of dashboard clicks, we use a Poisson quasi-maximum-likelihood (quasipoisson) regression to estimate Equations~\ref{eq:without-controls-qp} and~\ref{eq:without-controls-qp}. We use a quasi-maximum-likelihood model in order to account for potential overdispersion in the engagement data \citep{wooldridge_quasi-likelihood_1999}. 

\textbf{Without Controls}
\begin{align}\label{eq:without-controls-qp}
    log(Y_i) \sim \beta_0 &+ \beta_1 \text{incent}_i + \beta_2 \text{likes}_i + \beta_3 \text{dwell}_i  + \beta_4 \text{incent}_i \times \text{likes}_i + \beta_5 \text{incent}_i \times \text{dwell}_i + \epsilon_i
\end{align}

\textbf{With Warmup Session Controls}
\begin{align}\label{eq:with-controls-qp}
    log(Y_i) \sim  \beta_0 &+ \beta_1 \text{incent}_i + \beta_2 \text{likes}_i + \beta_3 \text{dwell}_i + \beta_4 \text{incent}_i \times \text{likes}_i + \beta_5 \text{incent}_i \times \text{dwell}_i + Y^0_i + \epsilon_i
\end{align}

\paragraph{OLS Models}\label{sec:model-ols}

For our two continuous variables i) Average song dwell time (logged) and ii) standard deviation song dwell time (logged), we use OLS regression to estimate Equations~\ref{eq:without-controls-ols} and~\ref{eq:with-controls-ols}.

\textbf{Without Controls}\
\begin{align}\label{eq:without-controls-ols}
    Y_i \sim \beta_0 &+ \beta_1 \text{incent}_i + \beta_2 \text{likes}_i + \beta_3 \text{dwell}_i  + \beta_4 \text{incent}_i \times \text{likes}_i + \beta_5 \text{incent}_i \times \text{dwell}_i + \epsilon_i
\end{align}

\textbf{With Warmup Session Controls}
\begin{align}\label{eq:with-controls-ols}
    Y_i \sim  \beta_0 &+ \beta_1 \text{incent}_i + \beta_2 \text{likes}_i + \beta_3 \text{dwell}_i + \beta_4 \text{incent}_i \times \text{likes}_i + \beta_5 \text{incent}_i \times \text{dwell}_i + Y^0_i + \epsilon_i
\end{align}

\subsection{Model, Proportion Regressions}\label{sec:model-proportion}
In addition to the outcomes reported in the main text, we also pre-registered two outcome variables: Proportion of Likes + Dislikes (per song listened) and Proportion of ``fast skips''  (per song listened). To estimate this, we estimate the following models using a quasi-binomial rate regression, where $Y_i$ is the count of our outcome of interest (either (i) Likes + Dislikes or (ii) ``fast skips'' , respectively) and $S_i$ is the total number of songs for user $i$. The results are shown in Table~\ref{tab:proportionTRUE}. 

\textbf{Without Controls}
\begin{align}\label{eq:without-controls-ols-prop}
    \log{\frac{Y_i}{S_i}}\sim  \beta_0 &+ \beta_1 \text{incent}_i + \beta_2 \text{likes}_i + \beta_3 \text{dwell}_i  + \beta_4 \text{incent}_i \times \text{likes}_i + \beta_5 \text{incent}_i \times \text{dwell}_i + \epsilon_i
\end{align}

\textbf{With Warm-up Session Controls}
\begin{align}\label{eq:with-controls-ols-prop}
    \log{\frac{Y_i}{S_i}} \sim  \beta_0 &+ \beta_1 \text{incent}_i + \beta_2 \text{likes}_i + \beta_3 \text{dwell}_i + \beta_4 \text{incent}_i \times \text{likes}_i + \beta_5 \text{incent}_i \times \text{dwell}_i + Y^0_i + \epsilon_i
\end{align}

\subsection{Regression Tables, Average Treatment Effect}\label{appendix:additional-table-ate}

Below we show the regression tables for the Average Treatment Effects (ATEs) for our outcome variables. Table~\ref{tab:quasi-poissonTRUE} shows the results for our count variables (i) Likes + Dislikes, (ii) ``fast skips'' , and (iii) Dashboard Clicks (from model equations specified in Section~\ref{sec:model-count}. Table~\ref{tab:olsTRUE} shows the results for our continuous dwell time variables (i) Log Dwell Time and (ii) Log St. Dev. Dwell Time (from model equations specified in Section~\ref{sec:model-ols}. Table~\ref{tab:proportionTRUE} shows the results for our proportion variables (i) Proportion of Likes + Dislikes and (ii) Proportion of ``fast skips'' from model~\ref{sec:model-proportion}.

\begin{table}[htbp]
   \caption{\label{tab:quasi-poissonTRUE} Quasi-Poisson Regression}
   \centering
   \small
   \begin{tabular}{lcccccc}
      \tabularnewline \midrule \midrule
      Dependent Var & \multicolumn{2}{c}{Likes + Dislikes} & \multicolumn{2}{c}{Fast Skips (5 sec)} & \multicolumn{2}{c}{Dashboard Clicks} \\ 
      Model:                               & (1)           & (2)           & (3)          & (4)          & (5)         & (6)\\  
      \midrule
      \emph{Variables}\\
      Constant                             & 2.1$^{***}$   & 1.6$^{***}$   & 1.7$^{***}$  & 1.2$^{***}$  & 3.0$^{***}$ & 2.4$^{***}$\\   
                                           & (0.05)        & (0.07)        & (0.10)       & (0.08)       & (0.07)      & (0.07)\\   
      1(Incentive)                         & 0.39$^{***}$  & 0.25$^{***}$  & 0.51$^{***}$ & 0.19         & 0.33$^{**}$ & 0.03\\   
                                           & (0.09)        & (0.07)        & (0.19)       & (0.15)       & (0.15)      & (0.13)\\   
      1(Likes Info)                        & 0.24$^{***}$  & 0.32$^{***}$  & -0.24        & -0.11        & 0.13        & 0.18\\   
                                           & (0.09)        & (0.08)        & (0.18)       & (0.16)       & (0.16)      & (0.15)\\   
      1(Dwell Info)                        & -0.39$^{***}$ & -0.36$^{***}$ & -0.22        & -0.23        & -0.17       & -0.16\\   
                                           & (0.10)        & (0.09)        & (0.18)       & (0.15)       & (0.15)      & (0.11)\\   
      1(Incentive) $\times$ 1(Likes Info)  & -0.03         & -0.05         & 0.21         & -0.25        & 0.08        & -0.50$^{*}$\\   
                                           & (0.19)        & (0.17)        & (0.36)       & (0.32)       & (0.31)      & (0.26)\\   
      1(Incentive) $\times$ 1(Dwell Info)  & -0.05         & 0.005         & -0.16        & -0.43        & -0.15       & -0.35\\   
                                           & (0.20)        & (0.18)        & (0.36)       & (0.30)       & (0.30)      & (0.22)\\   
      Likes + Dislikes, Warmup             &               & 0.05$^{***}$  &              &              &             &   \\   
                                           &               & (0.005)       &              &              &             &   \\   
      Fast Skips (5 sec), Warmup           &               &               &              & 0.05$^{***}$ &             &   \\   
                                           &               &               &              & (0.003)      &             &   \\   
      Dashboard Clicks, Warmup             &               &               &              &              &             & 0.02$^{***}$\\   
                                           &               &               &              &              &             & (0.001)\\   
      \midrule
      \emph{Fit statistics}\\
      Observations                         & 660           & 660           & 660          & 660          & 660         & 660\\  
      Squared Correlation                  & 0.09          & 0.40          & 0.02         & 0.35         & 0.02        & 0.36\\  
      \midrule \midrule
      \multicolumn{7}{l}{\emph{Heteroskedasticity-robust standard-errors in parentheses}}\\
      \multicolumn{7}{l}{\emph{Signif. Codes: ***: 0.01, **: 0.05, *: 0.1}}\\
   \end{tabular}
\end{table}

\begin{table}[htbp]
   \caption{\label{tab:olsTRUE} OLS Regression}
   \centering
   \small
   \begin{tabular}{lcccc}
      \tabularnewline \midrule \midrule
      Dependent Var & \multicolumn{2}{c}{Log Dwell Time} & \multicolumn{2}{c}{Log Std Dev Dwell Time} \\ 
      Model:                               & (1)           & (2)          & (3)           & (4)\\  
      \midrule
      \emph{Variables}\\
      Constant                             & 10.1$^{***}$  & 2.5$^{***}$  & 9.9$^{***}$   & 4.9$^{***}$\\   
                                           & (0.04)        & (0.36)       & (0.04)        & (0.46)\\   
      1(Incentive)                         & -0.27$^{***}$ & -0.02        & -0.33$^{***}$ & -0.14$^{*}$\\   
                                           & (0.08)        & (0.06)       & (0.08)        & (0.07)\\   
      1(Likes Info)                        & -0.15         & -0.04        & -0.31$^{***}$ & -0.18$^{**}$\\   
                                           & (0.09)        & (0.06)       & (0.09)        & (0.07)\\   
      1(Dwell Info)                        & -0.04         & 0.08         & 0.05          & 0.16$^{**}$\\   
                                           & (0.09)        & (0.06)       & (0.08)        & (0.07)\\   
      1(Incentive) $\times$ 1(Likes Info)  & -0.25         & 0.09         & -0.17         & 0.11\\   
                                           & (0.19)        & (0.12)       & (0.18)        & (0.15)\\   
      1(Incentive) $\times$ 1(Dwell Info)  & 0.09          & 0.21$^{*}$   & -0.02         & 0.10\\   
                                           & (0.18)        & (0.12)       & (0.17)        & (0.14)\\   
      Log Dwell Time, Warmup               &               & 0.74$^{***}$ &               &   \\   
                                           &               & (0.04)       &               &   \\   
      Log Std Dev Dwell Time, Warmup       &               &              &               & 0.50$^{***}$\\   
                                           &               &              &               & (0.05)\\   
      \midrule
      \emph{Fit statistics}\\
      Observations                         & 660           & 660          & 660           & 654\\  
      R$^2$                                & 0.02          & 0.56         & 0.05          & 0.30\\  
      Adjusted R$^2$                       & 0.02          & 0.56         & 0.04          & 0.30\\  
      \midrule \midrule
      \multicolumn{5}{l}{\emph{Heteroskedasticity-robust standard-errors in parentheses}}\\
      \multicolumn{5}{l}{\emph{Signif. Codes: ***: 0.01, **: 0.05, *: 0.1}}\\
   \end{tabular}
\end{table}

\begin{table}[htbp]
   \caption{\label{tab:proportionTRUE} Proportion Regression}
   \centering
   \small
   \begin{tabular}{lcccc}
      \tabularnewline \midrule \midrule
      Dependent Var & \multicolumn{2}{c}{Proportion of Likes + Dislikes} & \multicolumn{2}{c}{Proportion of Fast Skips (5 sec)} \\ 
      Model:                                   & (1)           & (2)           & (3)           & (4)\\  
      \midrule
      \emph{Variables}\\
      Constant                                 & 0.09          & -2.0$^{***}$  & -0.72$^{***}$ & -2.0$^{***}$\\   
                                               & (0.09)        & (0.14)        & (0.10)        & (0.10)\\   
      1(Incentive)                             & 0.17          & 0.25$^{*}$    & 0.33$^{*}$    & 0.11\\   
                                               & (0.17)        & (0.13)        & (0.19)        & (0.14)\\   
      1(Likes Info)                            & 0.57$^{***}$  & 0.78$^{***}$  & -0.37$^{**}$  & -0.31$^{**}$\\   
                                               & (0.17)        & (0.14)        & (0.18)        & (0.14)\\   
      1(Dwell Info)                            & -0.51$^{***}$ & -0.43$^{***}$ & -0.17         & -0.22$^{*}$\\   
                                               & (0.17)        & (0.15)        & (0.18)        & (0.13)\\   
      1(Incentive) $\times$ 1(Likes Info)      & -0.26         & -0.03         & 0.12          & 0.01\\   
                                               & (0.34)        & (0.30)        & (0.36)        & (0.29)\\   
      1(Incentive) $\times$ 1(Dwell Info)      & 0.06          & 0.06          & -0.08         & -0.14\\   
                                               & (0.34)        & (0.29)        & (0.37)        & (0.26)\\   
      Proportion of Likes + Dislikes, Warmup   &               & 3.8$^{***}$   &               &   \\   
                                               &               & (0.22)        &               &   \\   
      Proportion of Fast Skips (5 sec), Warmup &               &               &               & 4.4$^{***}$\\   
                                               &               &               &               & (0.21)\\   
      \midrule
      \emph{Fit statistics}\\
      Observations                             & 660           & 660           & 660           & 660\\  
      Squared Correlation                      & 0.07          & 0.50          & 0.01          & 0.59\\  
      \midrule \midrule
      \multicolumn{5}{l}{\emph{Heteroskedasticity-robust standard-errors in parentheses}}\\
      \multicolumn{5}{l}{\emph{Signif. Codes: ***: 0.01, **: 0.05, *: 0.1}}\\
   \end{tabular}
\end{table}

\subsection{Subgroup Analysis}\label{appendix:subgroup}%

Figure~\ref{fig:incentive-effects-age} shows the CATEs for the Incentive condition on our count variables for participants above and below 35, respectively. In the``Treatment'' Incentive condition, younger participants produced substantially more ````fast skips'' '' than those in the Incentive ``Control'' condition. We find that participants below 35 in the had 5.64 more ````fast skips'' '' in the ``Treatment'' vs. ``Control'' Incentive conditions; whereas participants above 35 had only 1.29 more ````fast skips'' '' in the ``Treatment'' vs. ``Control'' Incentive conditions, for a significant difference-in-CATEs of 4.36 ($p=.03$). In contrast, in the  Incentive ``Treatment'' condition (vs. Incentive ``Control'') older participants have 4.4 more likes and dislikes (compared to 2.7 for younger participants), and 8.5 more Dashboard Clicks (compared to 3.25 for younger participants) -- although the difference in CATEs for these two measures is not significantly different from zero. These findings, although exploratory, suggest that dwell time might be more salient metric for younger users than for older users. For example, TikTok, which has a particularly young userbase, uses watch time as the most important metric for generating new video recommendations \cite{wsjstaffTikTokAlgorithmWSJ2021}. On the other hand, older users show effects of substantial magnitude when it comes to more explicit forms of feedback, e.g. liking and disliking behavior.

\begin{figure}[t]
    \centering
    \includegraphics[width = \textwidth]{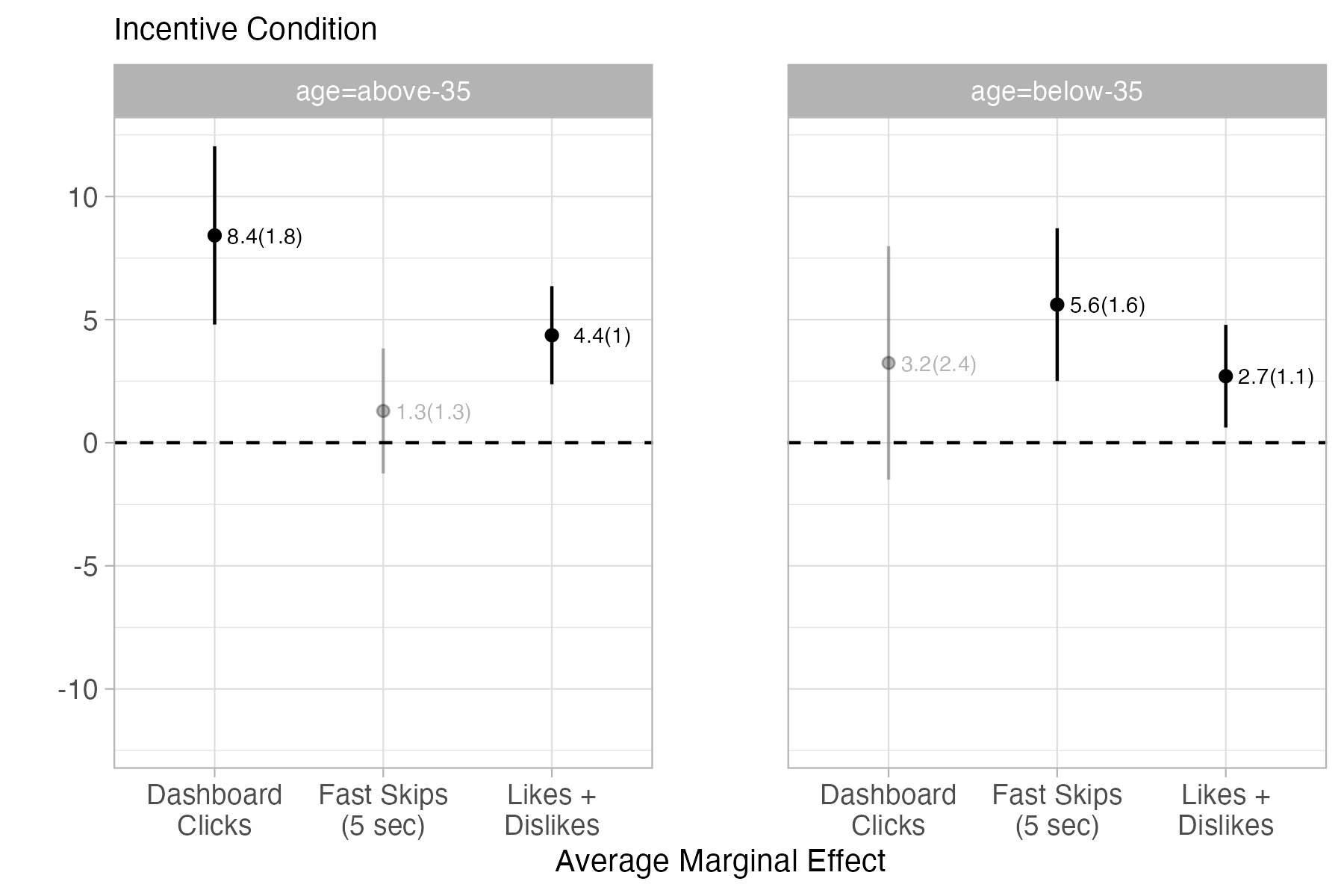}
    \caption{Conditional Average Treatment Effects (CATEs) for the Incentive condition  (i) Dashboard Clicks, (ii) ``fast skips'' , and (iii) Likes and Dislikes for the Incentive condition. Left: CATEs for participants above 35. Right: CATEs for participants below 35}
\label{fig:incentive-effects-age}
\end{figure}

In addition, we report additional subgroup analysis for (i) age (below or above 35) and (ii) Spotify use (``often'', coded as greater than once per week vs. ``rare'', coded as less than or equal to once per week) for our continuous dwell time variables.

We also examine self-reported behavior change (``yes'', coded as ``probably'' or ``definitely'' yes, vs. ``no'', coded as ``probably'' or ``definitely'' no) for our count variables and continuous dwell time variables as well. 

Interestingly, we do not observe evidence of substantial moderation by whether or not participants self-reported changing their behavior in our post-survey. For example, even participants who did not report changing their behavior showed significant increased numbers of likes and dislikes in the ``Likes Information'' Condition and fewer likes and dislikes in the ``Dwell Condition -- and the difference-in-CATEs between the two subgroups were not significant. This suggests that users might engage in ``strategization'' without consciously acknowledging that they are doing so. 

\begin{table}

\caption{\label{tab:dic_ols_incentive_condition_without_controls_age}Conditional Average Treatment Effects (CATEs) and Difference-in-CATEs (DIC) of the Incentive condition, by Age, Without Controls, pooled across Information Conditions}
\centering
\fontsize{9}{11}\selectfont
\begin{tabular}[t]{llll}
\toprule
\multicolumn{1}{c}{\bgroup\fontsize{10}{12}\selectfont \textbf{Outcome}\egroup{}} & \multicolumn{2}{c}{\bgroup\fontsize{10}{12}\selectfont \textbf{ATE}\egroup{}} & \multicolumn{1}{c}{\bgroup\fontsize{10}{12}\selectfont \textbf{DIC}\egroup{}} \\
\cmidrule{2-3}
\multicolumn{1}{c}{} & \multicolumn{1}{c}{Above-35} & \multicolumn{1}{c}{Below-35} & \multicolumn{1}{c}{}\\
\midrule
\multicolumn{1}{c}{Log Dwell Time} & \multicolumn{1}{c}{-0.21*(0.09)} & \multicolumn{1}{c}{-0.31**(0.12)} & \multicolumn{1}{c}{-0.1(0.15)}\\
\multicolumn{1}{c}{Log Std Dev Dwell Time} & \multicolumn{1}{c}{-0.38***(0.09)} & \multicolumn{1}{c}{-0.19†(0.11)} & \multicolumn{1}{c}{0.19(0.14)}\\
\bottomrule
\multicolumn{4}{l}{\textsuperscript{a} Heteroskedasticity Robust Standard Errors in Parentheses.}\\
\multicolumn{4}{l}{\textsuperscript{b} Signif. Codes: ***: .001, **: .01, *: .05, †: .1}\\
\end{tabular}
\end{table}

\begin{table}

\caption{\label{tab:dic_ols_information_condition_with_controls_age}Conditional Average Treatment Effects (CATEs) and Difference-in-CATEs (DIC) of the Information condition, by Age, With Controls, pooled across Incentive conditions}
\centering
\fontsize{9}{11}\selectfont
\begin{tabular}[t]{lllll}
\toprule
\multicolumn{1}{c}{\bgroup\fontsize{10}{12}\selectfont \textbf{Outcome}\egroup{}} & \multicolumn{1}{c}{\bgroup\fontsize{10}{12}\selectfont \textbf{Info condition}\egroup{}} & \multicolumn{2}{c}{\bgroup\fontsize{10}{12}\selectfont \textbf{ATE}\egroup{}} & \multicolumn{1}{c}{\bgroup\fontsize{10}{12}\selectfont \textbf{DIC}\egroup{}} \\
\cmidrule{3-4}
\multicolumn{1}{c}{} & \multicolumn{1}{c}{} & \multicolumn{1}{c}{Above-35} & \multicolumn{1}{c}{Below-35} & \multicolumn{1}{c}{}\\
\midrule
\multicolumn{1}{c}{Log Dwell Time} & \multicolumn{1}{c}{Likes} & \multicolumn{1}{c}{-0.06(0.07)} & \multicolumn{1}{c}{0.02(0.1)} & \multicolumn{1}{c}{0.08(0.12)}\\
\multicolumn{1}{c}{Log Dwell Time} & \multicolumn{1}{c}{Dwell} & \multicolumn{1}{c}{0.07(0.07)} & \multicolumn{1}{c}{0.08(0.1)} & \multicolumn{1}{c}{0.01(0.12)}\\
\multicolumn{1}{c}{Log Std Dev Dwell Time} & \multicolumn{1}{c}{Likes} & \multicolumn{1}{c}{-0.18†(0.09)} & \multicolumn{1}{c}{-0.15(0.12)} & \multicolumn{1}{c}{0.03(0.15)}\\
\multicolumn{1}{c}{Log Std Dev Dwell Time} & \multicolumn{1}{c}{Dwell} & \multicolumn{1}{c}{0.2*(0.09)} & \multicolumn{1}{c}{0.14(0.12)} & \multicolumn{1}{c}{-0.07(0.14)}\\
\bottomrule
\multicolumn{5}{l}{\textsuperscript{a} Heteroskedasticity Robust Standard Errors in Parentheses.}\\
\multicolumn{5}{l}{\textsuperscript{b} Signif. Codes: ***: .001, **: .01, *: .05, †: .1}\\
\end{tabular}
\end{table}

\begin{table}

\caption{\label{tab:dic_ols_incentive_condition_without_controls_spotify_how_often}Conditional Average Treatment Effects (CATEs) and Difference-in-CATEs (DIC) of the Incentive condition, by Spotify Use, Without Controls, pooled across Information Conditions}
\centering
\fontsize{9}{11}\selectfont
\begin{tabular}[t]{llll}
\toprule
\multicolumn{1}{c}{\bgroup\fontsize{10}{12}\selectfont \textbf{Outcome}\egroup{}} & \multicolumn{2}{c}{\bgroup\fontsize{10}{12}\selectfont \textbf{ATE}\egroup{}} & \multicolumn{1}{c}{\bgroup\fontsize{10}{12}\selectfont \textbf{DIC}\egroup{}} \\
\cmidrule{2-3}
\multicolumn{1}{c}{} & \multicolumn{1}{c}{Spotify Use=Often} & \multicolumn{1}{c}{Spotify Use=Rare} & \multicolumn{1}{c}{}\\
\midrule
\multicolumn{1}{c}{Log Dwell Time} & \multicolumn{1}{c}{-0.3***(0.09)} & \multicolumn{1}{c}{-0.17(0.11)} & \multicolumn{1}{c}{-0.13(0.15)}\\
\multicolumn{1}{c}{Log Std Dev Dwell Time} & \multicolumn{1}{c}{-0.3***(0.09)} & \multicolumn{1}{c}{-0.37**(0.12)} & \multicolumn{1}{c}{0.08(0.15)}\\
\bottomrule
\multicolumn{4}{l}{\textsuperscript{a} Heteroskedasticity Robust Standard Errors in Parentheses.}\\
\multicolumn{4}{l}{\textsuperscript{b} Signif. Codes: ***: .001, **: .01, *: .05, †: .1}\\
\end{tabular}
\end{table}

\begin{table}

\caption{\label{tab:dic_ols_information_condition_with_controls_spotify_how_often}Conditional Average Treatment Effects (CATEs) and Difference-in-CATEs (DIC) of the Information condition, by Spotify Use, With Controls, pooled across Incentive conditions}
\centering
\fontsize{9}{11}\selectfont
\begin{tabular}[t]{lllll}
\toprule
\multicolumn{1}{c}{\bgroup\fontsize{10}{12}\selectfont \textbf{Outcome}\egroup{}} & \multicolumn{1}{c}{\bgroup\fontsize{10}{12}\selectfont \textbf{Info condition}\egroup{}} & \multicolumn{2}{c}{\bgroup\fontsize{10}{12}\selectfont \textbf{ATE}\egroup{}} & \multicolumn{1}{c}{\bgroup\fontsize{10}{12}\selectfont \textbf{DIC}\egroup{}} \\
\cmidrule{3-4}
\multicolumn{1}{c}{} & \multicolumn{1}{c}{} & \multicolumn{1}{c}{Spotify Use=Often} & \multicolumn{1}{c}{Spotify Use=Rare} & \multicolumn{1}{c}{}\\
\midrule
\multicolumn{1}{c}{Log Dwell Time} & \multicolumn{1}{c}{Likes} & \multicolumn{1}{c}{-0.05(0.07)} & \multicolumn{1}{c}{0.03(0.1)} & \multicolumn{1}{c}{-0.08(0.12)}\\
\multicolumn{1}{c}{Log Dwell Time} & \multicolumn{1}{c}{Dwell} & \multicolumn{1}{c}{0.05(0.08)} & \multicolumn{1}{c}{0.13(0.09)} & \multicolumn{1}{c}{-0.08(0.12)}\\
\multicolumn{1}{c}{Log Std Dev Dwell Time} & \multicolumn{1}{c}{Likes} & \multicolumn{1}{c}{-0.24**(0.09)} & \multicolumn{1}{c}{0.04(0.13)} & \multicolumn{1}{c}{-0.28†(0.16)}\\
\multicolumn{1}{c}{Log Std Dev Dwell Time} & \multicolumn{1}{c}{Dwell} & \multicolumn{1}{c}{0.11(0.09)} & \multicolumn{1}{c}{0.35**(0.12)} & \multicolumn{1}{c}{-0.24(0.15)}\\
\bottomrule
\multicolumn{5}{l}{\textsuperscript{a} Heteroskedasticity Robust Standard Errors in Parentheses.}\\
\multicolumn{5}{l}{\textsuperscript{b} Signif. Codes: ***: .001, **: .01, *: .05, †: .1}\\
\end{tabular}
\end{table}

\begin{table}

\caption{\label{tab:dic_quasi-poisson_incentive_condition_without_controls_interaction_change}Conditional Average Treatment Effects (CATEs) and Difference-in-CATEs (DIC) of the Incentive condition, by Changed Interaction, Without Controls, pooled across Information Conditions}
\centering
\fontsize{9}{11}\selectfont
\begin{tabular}[t]{llll}
\toprule
\multicolumn{1}{c}{\bgroup\fontsize{10}{12}\selectfont \textbf{Outcome}\egroup{}} & \multicolumn{2}{c}{\bgroup\fontsize{10}{12}\selectfont \textbf{ATE}\egroup{}} & \multicolumn{1}{c}{\bgroup\fontsize{10}{12}\selectfont \textbf{DIC}\egroup{}} \\
\cmidrule{2-3}
\multicolumn{1}{c}{} & \multicolumn{1}{c}{No} & \multicolumn{1}{c}{Yes} & \multicolumn{1}{c}{}\\
\midrule
\multicolumn{1}{c}{Likes + Dislikes} & \multicolumn{1}{c}{4.26***(1.25)} & \multicolumn{1}{c}{2.86***(0.87)} & \multicolumn{1}{c}{1.4(1.52)}\\
\multicolumn{1}{c}{Fast Skips (5 sec)} & \multicolumn{1}{c}{4.01*(1.66)} & \multicolumn{1}{c}{2.22†(1.22)} & \multicolumn{1}{c}{1.79(2.06)}\\
\multicolumn{1}{c}{Dashboard Clicks} & \multicolumn{1}{c}{8.82*(4.07)} & \multicolumn{1}{c}{5.49(3.51)} & \multicolumn{1}{c}{3.33(5.37)}\\
\bottomrule
\multicolumn{4}{l}{\textsuperscript{a} Heteroskedasticity Robust Standard Errors in Parentheses.}\\
\multicolumn{4}{l}{\textsuperscript{b} Signif. Codes: ***: .001, **: .01, *: .05, †: .1}\\
\end{tabular}
\end{table}

\begin{table}

\caption{\label{tab:dic_quasi-poisson_information_condition_with_controls_interaction_change}Conditional Average Treatment Effects (CATEs) and Difference-in-CATEs (DIC) of the Information condition, by Changed Interaction, With Controls, pooled across Incentive conditions}
\centering
\fontsize{9}{11}\selectfont
\begin{tabular}[t]{lllll}
\toprule
\multicolumn{1}{c}{\bgroup\fontsize{10}{12}\selectfont \textbf{Outcome}\egroup{}} & \multicolumn{1}{c}{\bgroup\fontsize{10}{12}\selectfont \textbf{Info condition}\egroup{}} & \multicolumn{2}{c}{\bgroup\fontsize{10}{12}\selectfont \textbf{ATE}\egroup{}} & \multicolumn{1}{c}{\bgroup\fontsize{10}{12}\selectfont \textbf{DIC}\egroup{}} \\
\cmidrule{3-4}
\multicolumn{1}{c}{} & \multicolumn{1}{c}{} & \multicolumn{1}{c}{No} & \multicolumn{1}{c}{Yes} & \multicolumn{1}{c}{}\\
\midrule
\multicolumn{1}{c}{Likes + Dislikes} & \multicolumn{1}{c}{Likes} & \multicolumn{1}{c}{2.05*(0.96)} & \multicolumn{1}{c}{3.68***(1.01)} & \multicolumn{1}{c}{-1.63(1.42)}\\
\multicolumn{1}{c}{Likes + Dislikes} & \multicolumn{1}{c}{Dwell} & \multicolumn{1}{c}{-3.38**(1.05)} & \multicolumn{1}{c}{-2.7*(1.13)} & \multicolumn{1}{c}{-0.68(1.53)}\\
\multicolumn{1}{c}{Fast Skips (5 sec)} & \multicolumn{1}{c}{Likes} & \multicolumn{1}{c}{-0.55(1.48)} & \multicolumn{1}{c}{-0.34(1.29)} & \multicolumn{1}{c}{-0.21(1.97)}\\
\multicolumn{1}{c}{Fast Skips (5 sec)} & \multicolumn{1}{c}{Dwell} & \multicolumn{1}{c}{-1.7(1.35)} & \multicolumn{1}{c}{-0.53(1.23)} & \multicolumn{1}{c}{-1.17(1.82)}\\
\multicolumn{1}{c}{Dashboard Clicks} & \multicolumn{1}{c}{Likes} & \multicolumn{1}{c}{-1.94(5.26)} & \multicolumn{1}{c}{4.16(3.63)} & \multicolumn{1}{c}{-6.09(6.4)}\\
\multicolumn{1}{c}{Dashboard Clicks} & \multicolumn{1}{c}{Dwell} & \multicolumn{1}{c}{-3.83(3.39)} & \multicolumn{1}{c}{-4.65(3.57)} & \multicolumn{1}{c}{0.82(4.93)}\\
\bottomrule
\multicolumn{5}{l}{\textsuperscript{a} Heteroskedasticity Robust Standard Errors in Parentheses.}\\
\multicolumn{5}{l}{\textsuperscript{b} Signif. Codes: ***: .001, **: .01, *: .05, †: .1}\\
\end{tabular}
\end{table}

\begin{table}

\caption{\label{tab:dic_ols_incentive_condition_without_controls_interaction_change}Conditional Average Treatment Effects (CATEs) and Difference-in-CATEs (DIC) of the Incentive condition, by Changed Interaction, Without Controls, pooled across Information Conditions}
\centering
\fontsize{9}{11}\selectfont
\begin{tabular}[t]{llll}
\toprule
\multicolumn{1}{c}{\bgroup\fontsize{10}{12}\selectfont \textbf{Outcome}\egroup{}} & \multicolumn{2}{c}{\bgroup\fontsize{10}{12}\selectfont \textbf{ATE}\egroup{}} & \multicolumn{1}{c}{\bgroup\fontsize{10}{12}\selectfont \textbf{DIC}\egroup{}} \\
\cmidrule{2-3}
\multicolumn{1}{c}{} & \multicolumn{1}{c}{No} & \multicolumn{1}{c}{Yes} & \multicolumn{1}{c}{}\\
\midrule
\multicolumn{1}{c}{Log Dwell Time} & \multicolumn{1}{c}{-0.31**(0.1)} & \multicolumn{1}{c}{-0.18†(0.1)} & \multicolumn{1}{c}{-0.13(0.14)}\\
\multicolumn{1}{c}{Log Std Dev Dwell Time} & \multicolumn{1}{c}{-0.43***(0.1)} & \multicolumn{1}{c}{-0.2*(0.09)} & \multicolumn{1}{c}{-0.24†(0.14)}\\
\bottomrule
\multicolumn{4}{l}{\textsuperscript{a} Heteroskedasticity Robust Standard Errors in Parentheses.}\\
\multicolumn{4}{l}{\textsuperscript{b} Signif. Codes: ***: .001, **: .01, *: .05, †: .1}\\
\end{tabular}
\end{table}

\begin{table}

\caption{\label{tab:dic_ols_information_condition_with_controls_interaction_change}Conditional Average Treatment Effects (CATEs) and Difference-in-CATEs (DIC) of the Information condition, by Changed Interaction, With Controls, pooled across Incentive conditions}
\centering
\fontsize{9}{11}\selectfont
\begin{tabular}[t]{lllll}
\toprule
\multicolumn{1}{c}{\bgroup\fontsize{10}{12}\selectfont \textbf{Outcome}\egroup{}} & \multicolumn{1}{c}{\bgroup\fontsize{10}{12}\selectfont \textbf{Info condition}\egroup{}} & \multicolumn{2}{c}{\bgroup\fontsize{10}{12}\selectfont \textbf{ATE}\egroup{}} & \multicolumn{1}{c}{\bgroup\fontsize{10}{12}\selectfont \textbf{DIC}\egroup{}} \\
\cmidrule{3-4}
\multicolumn{1}{c}{} & \multicolumn{1}{c}{} & \multicolumn{1}{c}{No} & \multicolumn{1}{c}{Yes} & \multicolumn{1}{c}{}\\
\midrule
\multicolumn{1}{c}{Log Dwell Time} & \multicolumn{1}{c}{Likes} & \multicolumn{1}{c}{0.07(0.08)} & \multicolumn{1}{c}{-0.14(0.09)} & \multicolumn{1}{c}{0.21†(0.12)}\\
\multicolumn{1}{c}{Log Dwell Time} & \multicolumn{1}{c}{Dwell} & \multicolumn{1}{c}{0.11(0.08)} & \multicolumn{1}{c}{0(0.1)} & \multicolumn{1}{c}{0.12(0.12)}\\
\multicolumn{1}{c}{Log Std Dev Dwell Time} & \multicolumn{1}{c}{Likes} & \multicolumn{1}{c}{-0.04(0.12)} & \multicolumn{1}{c}{-0.27**(0.11)} & \multicolumn{1}{c}{0.24(0.16)}\\
\multicolumn{1}{c}{Log Std Dev Dwell Time} & \multicolumn{1}{c}{Dwell} & \multicolumn{1}{c}{0.14(0.1)} & \multicolumn{1}{c}{0.18†(0.1)} & \multicolumn{1}{c}{-0.04(0.14)}\\
\bottomrule
\multicolumn{5}{l}{\textsuperscript{a} Heteroskedasticity Robust Standard Errors in Parentheses.}\\
\multicolumn{5}{l}{\textsuperscript{b} Signif. Codes: ***: .001, **: .01, *: .05, †: .1}\\
\end{tabular}
\end{table}

\pagebreak

\subsection{Outcome Distributions}\label{appendix:ecdf}

While understanding the average treatment effect is important in light of our hypotheses, another relevant consideration is how evidence of strategization manifests at different values of our outcome distribution. Research has shown that engagement behavior on digital platforms is highly skewed. Are the effects we see driven by a few particularly highly active outliers, or do we see consistent evidence of strategic behavior across our outcome distributions? 

We shed light on this question by plotting the empirical cumulative-distribution functions (eCDFs) of each outcome by condition. This approach allows us to understand how our treatments affect behavior not just on average, but across all values of our outcomes. We note that these analyses were not pre-registered but done post-hoc. 

The  distributions across (i) outcome variables and (ii) conditions are visualized in Figures~\ref{fig:ecdf-qp}. The X-axis shows different values of the outcomes of interest, the Y-axis shows the proportion of users who fall at or below each value, with results plotted separately for the ``Treatment'' Incentive vs. ``Control'' Incentive conditions (left) and ``Likes Information'' vs. ``Dwell'' Information vs. ``Control'' Information conditions (Right). At each value $x$ of the outcome variable, we perform a difference-in-proportion test; values of $x$ where the proportion of users differs in the treatment condition significantly from the proportion in the respective control the $\alpha=.05$ level are shown in bold. This procedure allows us to understand how our distributions differ at a granular level, rather than testing for overall differences in distributions. 

For our count variables, we can see that the proportion of users with very little activity does not change between the ``Treatment'' Incentive and ``Control'' Incentive groups. For example, in the ``Treatment'' Incentive group, 15\% of users have one or fewer likes compared to 14\% in the ``Control'' Incentive group ($p=.8$), 53\% have one or fewer ``fast skip'' compared to 59\% in ``Control'' ($p=.11$) and 11\% of users had one or fewer Dashboard Clicks  -- the same as in the ``Control'' ($p=.97)$. However, at greater values of our outcome variables, we do observe differences. For example, we find that 52.4\% of users in the ``Control'' Incentive condition have 5 likes or fewer, compared to only 41.0\% in the ``Treatment'' Incentive condition ($p=.004$); 77\% have 5 ``fast skips''  or fewer compared to 68.2\% in Control ($p = .01$); and 41\% have 5 Dashboard Clicks or fewer compared to 30\% in Control ($p = .005)$. The proportion of users engaging at very high levels is also subsantially larger in the ``Treatment'' Incentive group. We see that 17\% of users in the have more than 20 ``likes'' and ``dislikes'' group compared to only 6\% in the ``Control'' Incentive group ($p < .001$); 12\% of users have greater than 20 ``fast skips''  compared to 6\% in Control ($p = .03$), and 30\% of users have 20 or more dashboard clicks in ``Treatment'' Incentive compared to  19\% in ``Control'' Incentive ($p=.0004)$. These results suggest that our ``Treatment'' Incentive condition increases activity among those who engage in the behavior in the first place -- particularly among those who are very active. On the other hand, those who eschew the behavior are not driven to take it up when incentives increase. 

In contrast, we see evidence that receiving new information about how the platform is learning can change the \textit{type} of behavior users engage in, not just the level of the behavior. For example, 15\% of participants in the ``Control'' Information group had one or fewer ``likes'' and ``dislikes,'' compared to just 4\% of participants in the ``Likes'' ($p=.0004$) and 23\% of users in the ``Dwell'' group ($p = .04$). This suggests that informing people that the platform is tracking ``likes'' and ``dislikes'' causes  participants to start engaging in like behavior when they otherwise would not. Correspondingly, implying that the platform is \textit{not} tracking likes (by calling attention to ``dwell time'') induces some participants to abandon the behavior entirely. Furthermore, a much smaller proportion of participants ``like'' at high rates in the ``Dwell'' condition -- just 3\% of users had greater than 20 ``likes'' and ``dislikes,'' compared to 13\% in the ``Control'' condition. On the other hand, and in contrast to what we observed in the ``Incentive'' condition, we do not see a significant difference in the proportion of participants engaging at very high rates in the ``Likes'' condition. One explanation for this is that participants who engaged at high rates might have already assumed the platform was tracking ``likes'' and ``dislikes,'' and therefore explicitly being informed of that fact had little additional effect. 

These findings show that evidence of strategization is not just concentrated among outliers or the most active participants, but apparent across our outcome distribution. Furthermore, we shed light on different types of strategic behavior  -- we see evidence that our treatments change not only the level of user feedback (particularly in the case of the ``Incentive'' condition), but also the \textit{type} of feedback (in the case of the ``Information'' condition. 

We now consider our continuous outcome variables -- Log Dwell Time and Log Standard Deviation of Dwell Time. In the Incentive condition, we see a relatively homogenous shift in the outcome distributions, showing that participants in the ``Treatment'' Incentive have a shorter average and smaller standard deviation of dwell time than those in the Incentive Control across different percentiles of our outcome variable. For example, we see that 20\% of users in the ``Treatment'' condition have an average dwell time less than 10 seconds compared to 11\% in the Control Incentive condition ($p = .004$), and 63\% of users have an average dwell time less than 30 seconds compared to 54\% in the Incentive Control condition ($p = .02$). Similarly, we see that 27\% of users had a standard deviation of dwell time of 10 seconds or less in the ``Treatment'' Incentive condition, compared to 15\% in Incentive Control ($p = .0003$) and 73\% had a standard deviation of dwell time of 30 seconds or less compared to 60\% in the Incentive Control condition ($p = .0004$). However, we see similar proportions of users have extremely high level of dwell times and standard deviations: 10\% of users in the Treatment Incentive condition and 12\% of users in the Control condition have average dwell times greater than 90 seconds ($p = .4$), and 4\% of users in the Treatment Incentive condition and 6\% of users in the Control Incentive condition had standard deviations greater than 90 seconds ($p = .5$).

\begin{figure}[h!]
    \centering
    \includegraphics[]{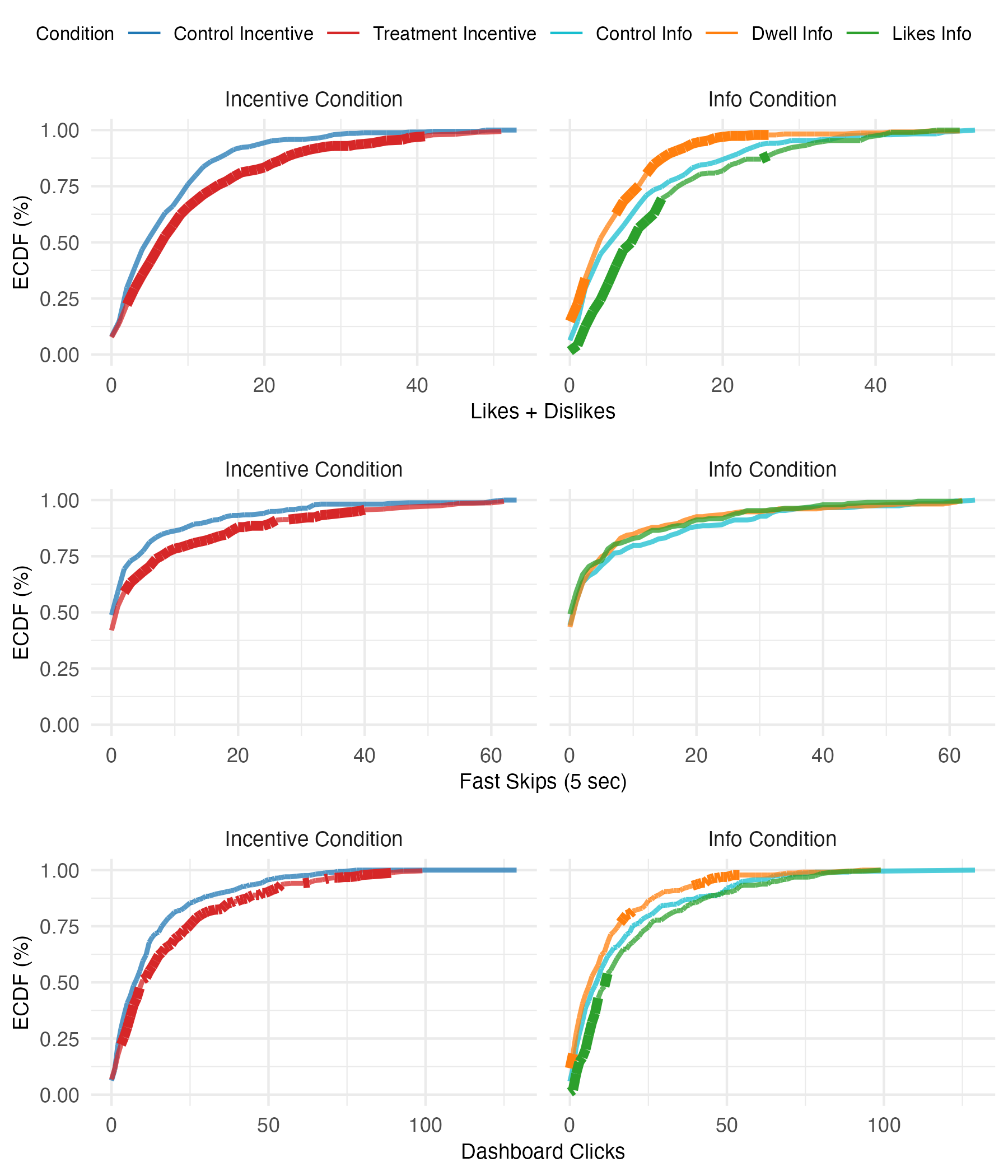}
    \caption{Empirical CDFs of Likes and Dislikes (Top), ``fast skips''  (Middle) and Dashboard clicks for the Incentive condition (Left) and Information condition (Right). Bold distributions signify that a difference-in-proportion test was significant at that value of the outcome variable at the $p < .05$ level.}
    \label{fig:ecdf-qp}
\end{figure}

\begin{figure}[h!]
    \centering
    \includegraphics[]{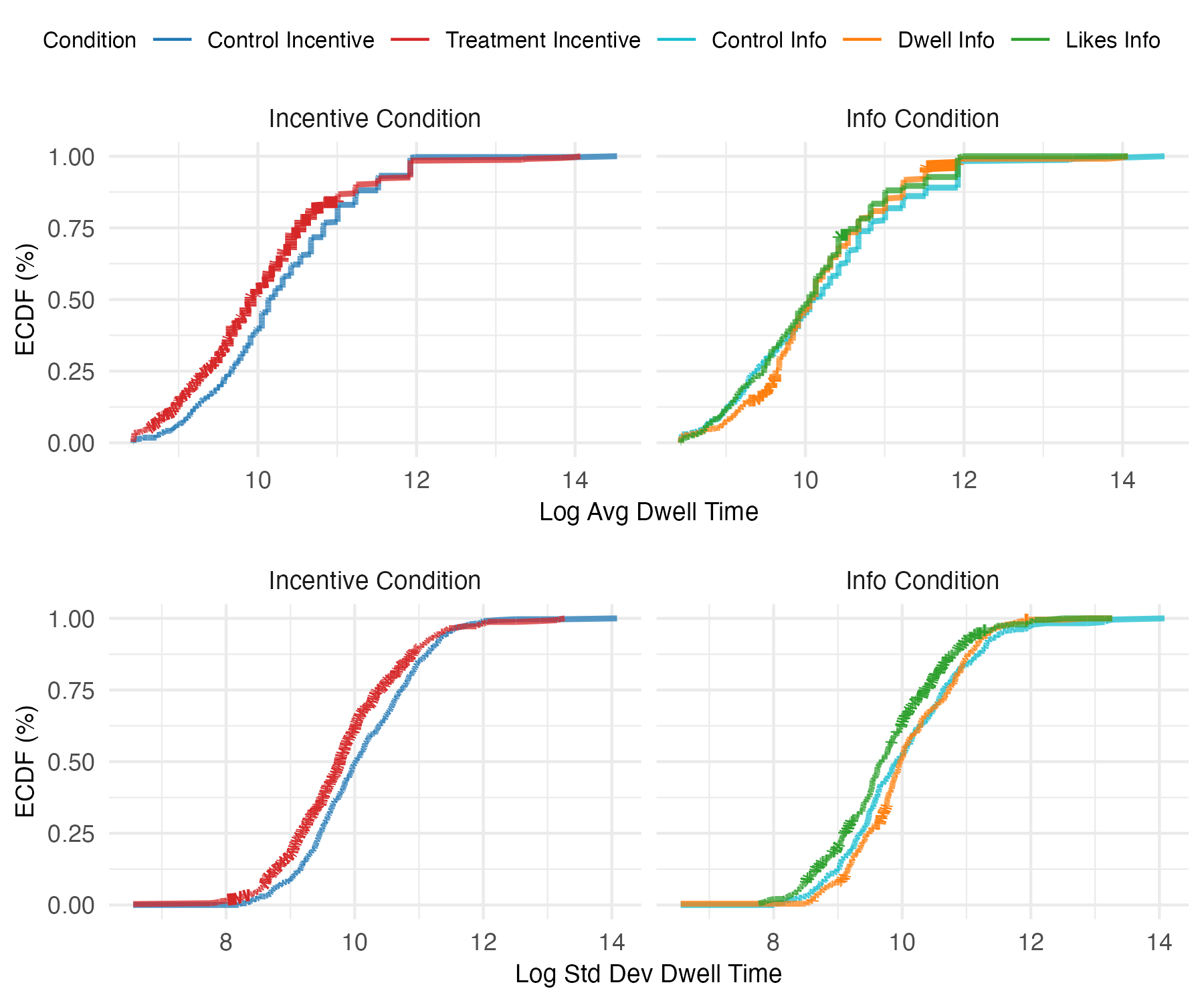}
    \caption{Empirical CDFs of Log Dwell Time (Top) and Log Standard Deviation of Dwell Time by Incentive condition (Left) and Information condition (Right). Bold distributions signify that a difference-in-proportion test was significant at that value of the outcome variable at the $p < .05$ level.}
    \label{fig:ecdf-ols}
\end{figure}

\clearpage
\section{Additional post-experiment survey results}\label{app:additional_survey_results}

\subsection{Post-Experiment Survey Questions}\label{app:survey_Qs}

\begin{enumerate}
    \item \emph{Did the way that you interacted with songs change across the three listening sessions?}
    Answers: (a) Definitely yes, (b) Probably yes, (c) Probably no, (d) Definitely no, (e) I don't know.

    \item \emph{If yes, how did your interactions change across listening sessions? (CHECK ALL THAT APPLY)} 
    Answers: (a) I changed how much I used the thumbs-up/down buttons, 
    (b) I changed how much I used the skip button, 
    (c) I changed how much I used the restart button, 
    (d) I changed how long I spent on each song, 
    (e) I'm not sure.

    \item \emph{How do you think platforms like Spotify choose what to show you on your homepage? (CHECK ALL THAT APPLY)}
    Answers: 
    (a) Based on what’s most popular across the platform, 
    (b) By randomly selecting songs you've recently listened to,
    (c) By analyzing what you've liked or skipped on the platform, 
    (d) By randomly selecting songs that editors have picked, 
    (e) Based on your age, gender, and location, 
    (f) I don’t know.

    \item \emph{How do you think social media platforms like Facebook, Twitter, or TikTok choose what to show you? (CHECK ALL THAT APPLY)}
    Answers: 
    (a) Based on what’s currently trending across the platform,
    (b) By randomly selecting recent posts on the platform, 
    (c) By analyzing what posts you've liked/commented on/etc., 
    (d) By analyzing how long you watch videos and how you scroll down your feed,
    (e) By randomly selecting posts that editors at the platform pick,
    (f) Based on your age, gender, and location, 
    (g) I don’t know.

    \item \emph{Do you ever try to “talk” to your algorithm or “hide” things from it? For example, do you ever give a song a “thumbs-up” just to Spotify that you want to see similar songs? Or do you sometimes avoid clicking on an advertisement just because you’re worried about getting many similar advertisements in the future? If you do, tell us how and why.}
    Participants are permitted to provide open-ended, text answers to this question.

    \item \emph{Are you concerned about data privacy online?}
    Answers: 
    (a) Yes, I'm very concerned,
    (b) I'm sometimes concerned,
    (c) I'm rarely concerned, 
    (d) No, I'm not concerned at all, 
    (e) I don't know what data privacy is.

    \item \emph{How often do you use music recommendation platforms, like Spotify?}
    Answers: 
    (a) A few hours everyday, 
    (b) A few hours each week, 
    (c) A few hours each month, 
    (d) Less than a few hours each month, 
    (e) Never. 

    \item \emph{How old are you?}
    Answers: 
    (a) 18-25, 
    (b) 25-35, 
    (c) 45-55, 
    (d) 55+.

    \item \emph{What is the highest level of education you have completed?}
    Answers: 
    (a) Some high school or less, 
    (b) High school diploma or GED, 
    (c) Some college but no degree,
    (d) Associates or technical degre, 
    (e) Bachelor's degree, 
    (f) Graduate or professional degree, 
    (g) Prefer not to say. 

    \item \emph{Any comments, questions, or feedback?}
    Participants are permitted to provide open-ended, text answers to this question.
\end{enumerate}
The order of the answers is randomized for Questions 2-4.

\clearpage
\subsection{Demographics}\label{app:demographic_plots}

The plots below summarize the demographic groups represented by our study. 
There are more plots showing the demographic split across different treatment groups (i.e., verify whether our randomization was effective) in the ``figures/post\_experiment\_plots'' folder.

\begin{figure}[!htbp]
    \centering
    \begin{subfigure}[b]{0.49\textwidth}
        \includegraphics[width=\textwidth]{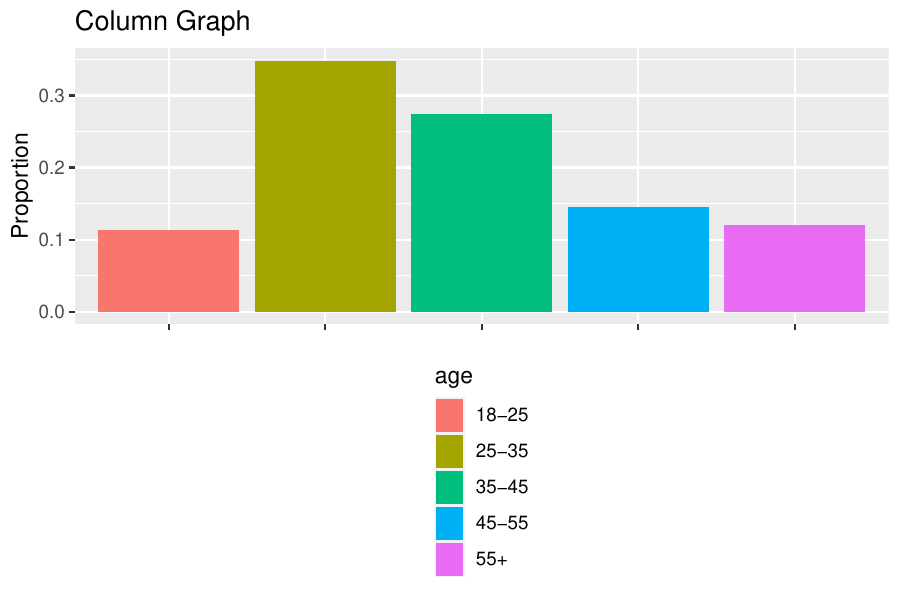}
        \caption{}
        \label{fig:12-sub1}
    \end{subfigure}
    \hfill
    \begin{subfigure}[b]{0.49\textwidth}
        \includegraphics[width=\textwidth]{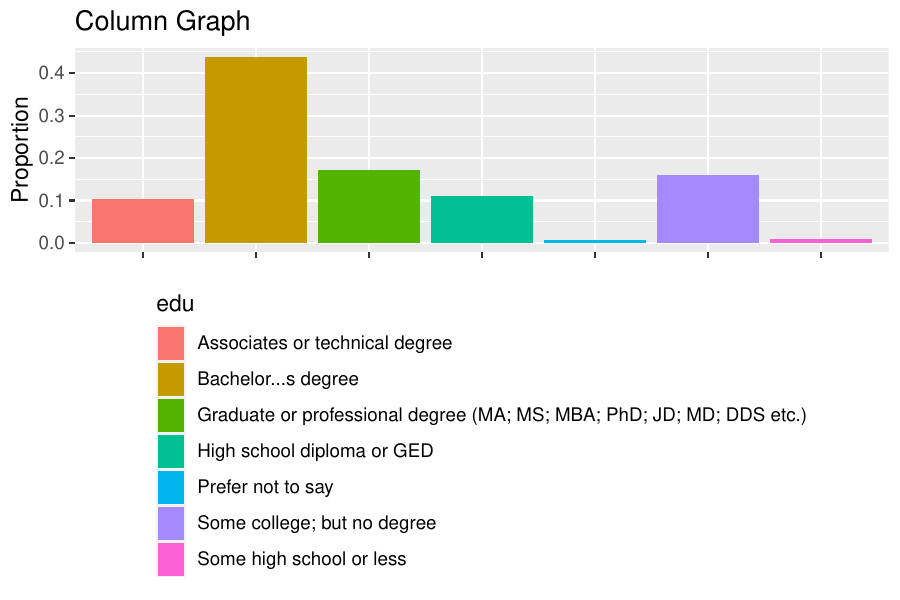}
        \caption{}
        \label{fig:12-sub2}
    \end{subfigure}
    \hfill
    \begin{subfigure}[b]{0.49\textwidth}
        \includegraphics[width=\textwidth]{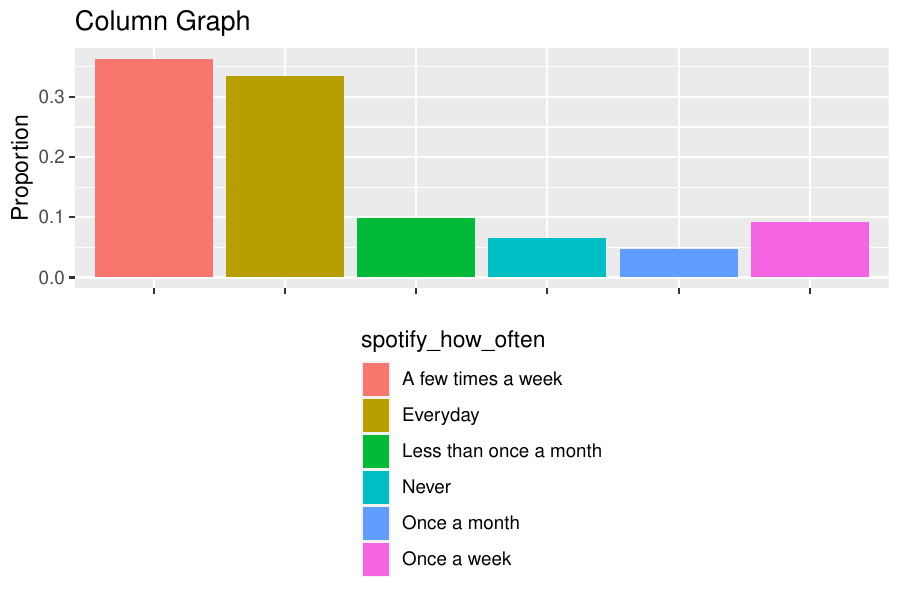}
        \caption{}
        \label{fig:12-sub3}
    \end{subfigure}
    \hfill
    \begin{subfigure}[b]{0.49\textwidth}
        \includegraphics[width=\textwidth]{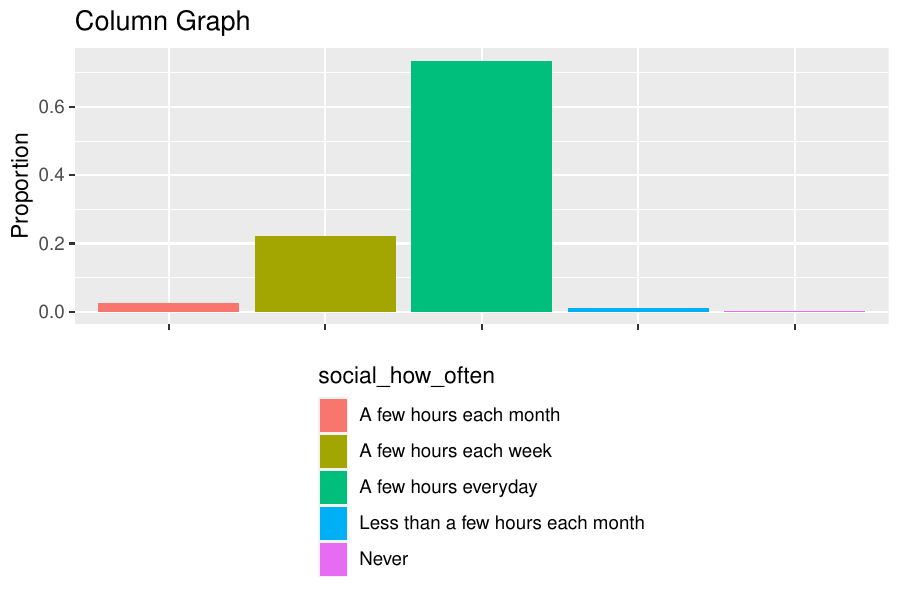}
        \caption{}
        \label{fig:12-sub4}
    \end{subfigure}
    \hfill
    \begin{subfigure}[b]{0.49\textwidth}
        \includegraphics[width=\textwidth]{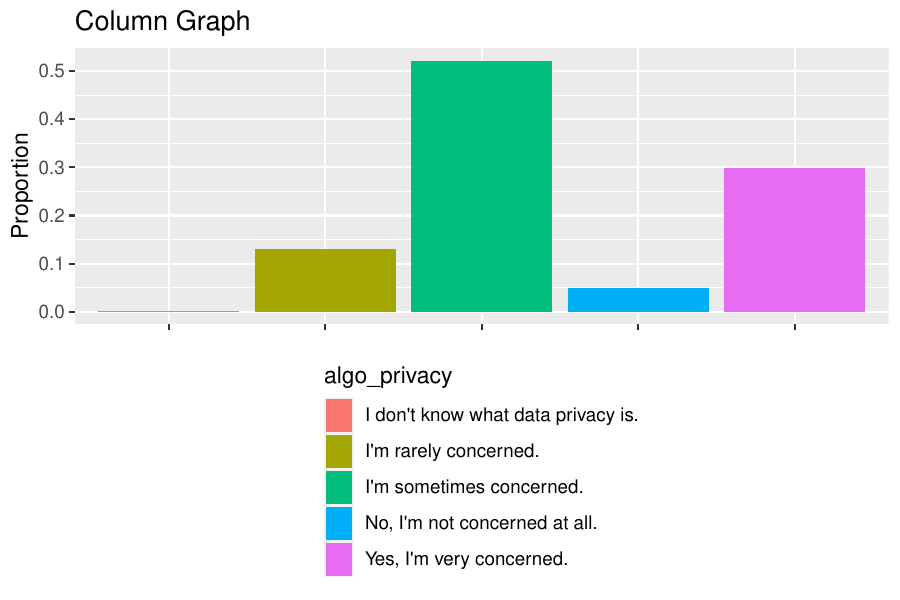}
        \caption{}
        \label{fig:12-sub5}
    \end{subfigure}
    \hfill
    \begin{subfigure}[b]{0.49\textwidth}
        \includegraphics[width=\textwidth]{figures/post_experiment_plots/age_.pdf}
        \caption{}
        \label{fig:12-sub6}
    \end{subfigure}
    \caption{Demographic responses}
\end{figure}

\clearpage

\subsection{How participants believe online algorithms work}

\begin{figure}[!htbp]
    \centering
    \begin{subfigure}[b]{0.49\textwidth}
        \includegraphics[width=\textwidth]{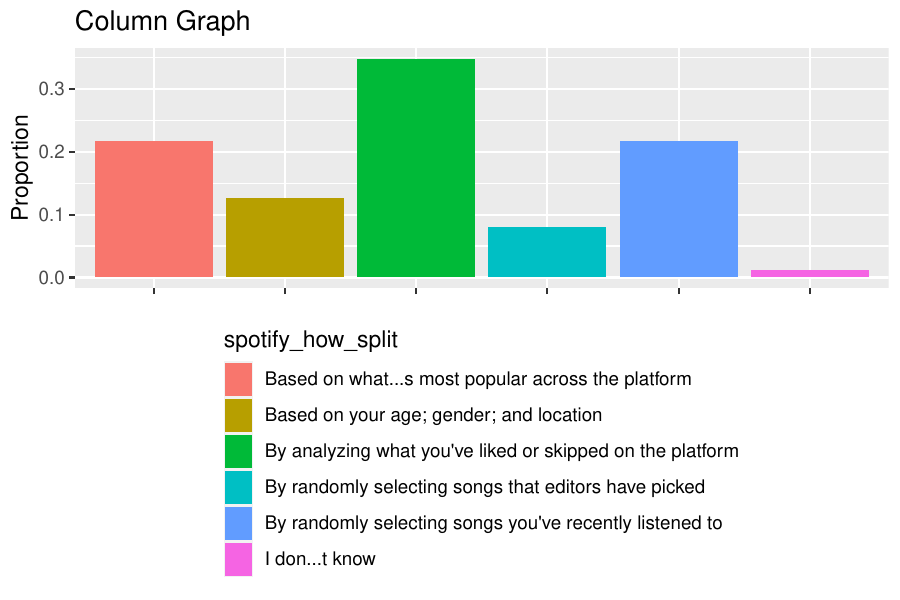}
        \caption{}
        \label{fig:13-sub1}
    \end{subfigure}
    \hfill
    \begin{subfigure}[b]{0.49\textwidth}
        \includegraphics[width=\textwidth]{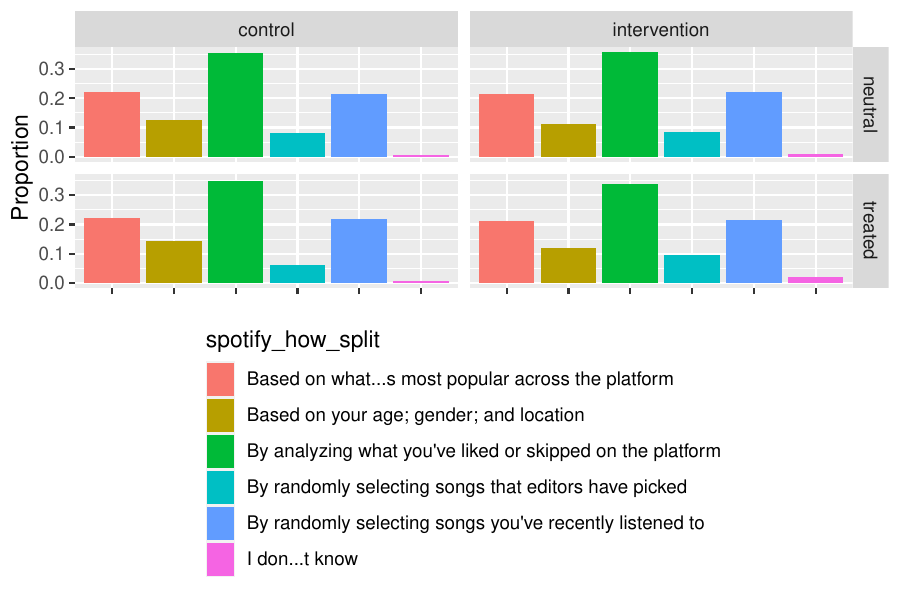}
        \caption{}
        \label{fig:13-sub2}
    \end{subfigure}
    \hfill
    \begin{subfigure}[b]{0.49\textwidth}
        \includegraphics[width=\textwidth]{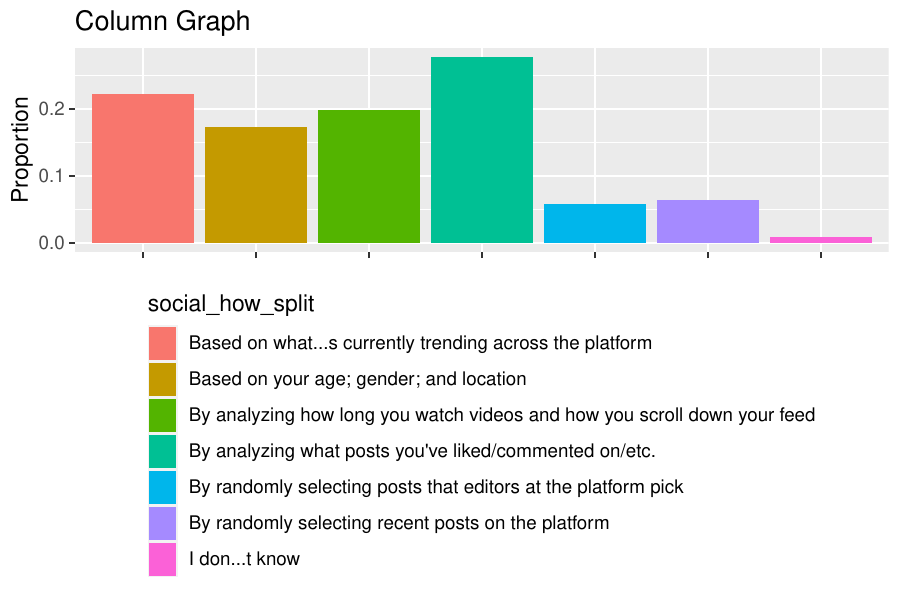}
        \caption{}
        \label{fig:13-sub3}
    \end{subfigure}
    \hfill
    \begin{subfigure}[b]{0.49\textwidth}
        \includegraphics[width=\textwidth]{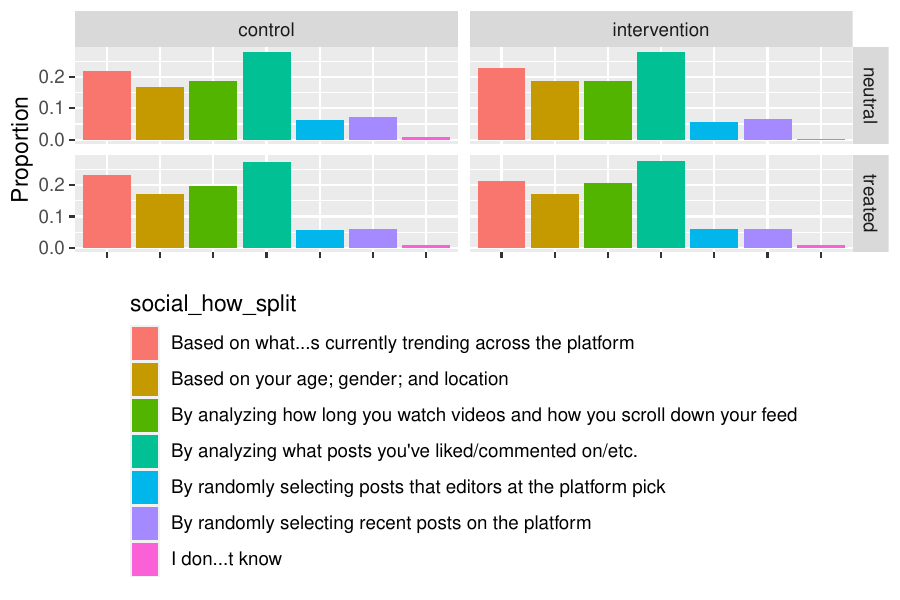}
        \caption{}
        \label{fig:13-sub4}
    \end{subfigure}
    \caption{How participants believe that Spotify (top) and social media algorithms (bottom) work.}
\end{figure}
\clearpage

\end{APPENDICES}

\ACKNOWLEDGMENT{The authors would like to thank Dean Eckles, Drazen Prelec, and David Rand for useful feedback. We would also like to thank the
members of the M\k{a}dry Lab for their role in piloting and reporting bugs in an earlier version of the custom music player. 
}

\end{document}